\pdfoutput=1
\newcommand*{\ATLASLATEXPATH}{./}
\documentclass[cernpreprint, english, texlive=2016]{\ATLASLATEXPATH atlasdoc}

\usepackage{\ATLASLATEXPATH atlaspackage}
\usepackage{rotating}
\usepackage{verbatim}
\usepackage{multirow}
\usepackage{bm}
\usepackage{\ATLASLATEXPATH atlasbiblatex}

\usepackage{\ATLASLATEXPATH atlasphysics}

\graphicspath{{./}}

\usepackage{ANA-EXOT-2017-31-PAPER-defs}

\AtlasTitle{Combination of searches for heavy resonances decaying into bosonic and leptonic final states using 36~fb$^{-1}$ of proton--proton collision data at $\sqrt{s} = 13$~TeV with the ATLAS detector}

\AtlasAbstract{%
Searches for new heavy resonances decaying into different pairings of $W$, $Z$, or Higgs bosons, as well as directly into leptons, are presented using a data sample corresponding to 36.1~\ifb\ of $pp$ collisions at $\sqrt{s}$ = 13~\TeV\ collected during 2015 and 2016 with the ATLAS detector at the CERN Large Hadron Collider. Analyses selecting bosonic decay modes in the $qqqq$, $\nu\nu qq$, $\ell\nu qq$, $\ell\ell qq$, $\ell\nu\ell\nu$, $\ell\ell\nu\nu$, $\ell\nu\ell\ell$, $\ell\ell\ell\ell$, $qqbb$, $\nu\nu bb$, $\ell\nu bb$, and $\ell\ell bb$ final states are combined, searching for a narrow-width resonance. Likewise, analyses selecting the leptonic $\ell\nu$ and $\ell\ell$ final states are also combined. These two sets of analyses are then further combined. No significant deviation from the Standard Model predictions is observed. Three benchmark models are tested: a model predicting the existence of a new heavy scalar singlet, a simplified model predicting a heavy vector-boson triplet, and a bulk Randall--Sundrum model with a heavy spin-2 Kaluza--Klein excitation of the graviton. Cross-section limits are set at the 95\% confidence level using an asymptotic approximation and are compared with predictions for the benchmark models. These limits are also expressed in terms of constraints on couplings of the heavy vector-boson triplet to quarks, leptons, and the Higgs boson. The data exclude a heavy vector-boson triplet with mass below 5.5~\TeV\ in a weakly coupled scenario and 4.5~\TeV\ in a strongly coupled scenario, as well as a Kaluza--Klein graviton with mass below 2.3~\TeV.}

\author{The ATLAS Collaboration}

\AtlasRefCode{ANA-EXOT-2017-31}

\PreprintIdNumber{CERN-EP-2018-179}

\AtlasJournalRef{Phys.\ Rev.\ D 98 (2018) 052008}
\AtlasDOI{10.1103/PhysRevD.98.052008}

\hypersetup{pdftitle={ATLAS document},pdfauthor={The ATLAS Collaboration}}

\begin{document}

\maketitle


\section{Introduction}
\label{sec:intro}
The search for new heavy particles is an important part of the physics program
at the Large Hadron Collider (LHC) and has been the focus of an intense effort
to uncover physics beyond the Standard Model (SM) in a broad range of final states.
Many of these searches are motivated by models aiming to resolve the hierarchy problem
such as the Randall--Sundrum (RS) model with a warped extra dimension~\cite{RSmodel},
by models with extended Higgs sectors as in the two-Higgs-doublet model~\cite{Branco:2011iw},
or by models with composite Higgs bosons~\cite{Contino:2011np} 
or extended gauge sectors as in Grand Unified 
Theories~\cite{Pati:1974yy,Georgi:1974sy,Fritzsch:1974nn}.

Although no significant excess has been observed to date, strong constraints
have been placed on the production of such new heavy particles.
A combination of searches for the production of heavy resonances decaying
into the $VV$ (with $V = W$ or $Z$) final state
in proton--proton ($pp$) collisions at a center-of-mass energy $\rts = 13$~\TeV\
corresponding to an integrated luminosity of 3.2~\ifb\ has been published
by the ATLAS Collaboration~\cite{EXOT-2016-01}.
Similarly, a combination of searches in the $VV$ and $VH$ (with $H$ representing the SM Higgs boson) final states obtained with 19.7~\ifb\ at $\rts = 8$ \TeV\ and
2.7~\ifb\ at $\rts = 13$ \TeV\ has been published by the CMS Collaboration~\cite{CMS-B2G-16-007}. In this article, the combination is broadened to include the results of not only the $VV$ and $VH$ searches but lepton–antilepton searches as well. It uses the most recent ATLAS results obtained at $\rts = 13$~\TeV\ with an integrated luminosity of approximately 36~\ifb. A combination with a broader set of final states allows one to explore the complementarity of these searches and set stronger constraints over a wider range of models of physics beyond the SM. Several diagrams illustrating the production and decay of new heavy resonances are shown in Figure~\ref{fig:FeynmanDiags}.

The specific searches combined in this article are those performed in 
the $VV$ channels:
$WZ \to qqqq$~\cite{EXOT-2016-19},
$\ell\nu qq$~\cite{EXOT-2016-28}, 
$\ell\nu\ell\ell$~\cite{EXOT-2016-11}, 
$WW \to qqqq$~\cite{EXOT-2016-19},
$\ell\nu qq$~\cite{EXOT-2016-28}, 
$\ell\nu\ell\nu$~\cite{HIGG-2016-31}, 
and
$ZZ \to qqqq$~\cite{EXOT-2016-19},
$\nu\nu qq$~\cite{EXOT-2016-29},
$\ell\ell qq$~\cite{EXOT-2016-29}, 
$\ell\ell\nu\nu$~\cite{HIGG-2016-19},
$\ell\ell\ell\ell$~\cite{HIGG-2016-19};
the $VH$ channels:
$WH \to qq bb$~\cite{EXOT-2016-12}, $\ell\nu bb$~\cite{EXOT-2016-10}, and
$ZH \to qq bb$~\cite{EXOT-2016-12}, $\nu\nu bb$~\cite{EXOT-2016-10}, 
$\ell\ell bb$~\cite{EXOT-2016-10};
and the lepton--antilepton channels: 
$\ell\nu$~\cite{EXOT-2016-06} and $\ell\ell$~\cite{EXOT-2016-05}.
The charged leptons $\ell$ are either electrons or muons. 
For the $VV$ and $VH$ decay channels involving leptonic decays of vector bosons,
$\tau$-leptons are included as part of the signal since $\tau$-lepton decays
into electrons or muons provide a small amount of additional acceptance.
The impact of $\tau$-leptons is very small and neglected in other channels.
In this article,
the $VV$ and $VH$ decay channels are collectively named ``bosonic'', whereas
the lepton--antilepton decay channels are collectively named ``leptonic.'' 
The analyses generally search for narrow resonances in the final-state mass distribution
with the signal shape extracted from Monte Carlo (MC) simulation of specific models.
The background shape and normalization are extracted from a combination of MC simulation
and data, often relying on dedicated control regions to extract the various background
contributions.
The mass distributions and associated systematic uncertainties from the various channels are combined taking correlations into account, as described below.

\begin{figure}[tb]
	\centering
	\subfloat[]{
      		\includegraphics[width=0.35\textwidth]{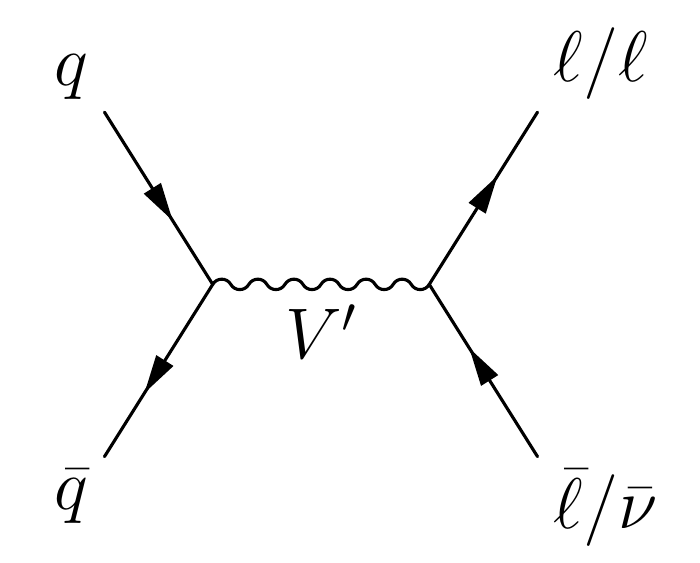}
      	}
      \subfloat[]{
      		\includegraphics[width=0.35\textwidth]{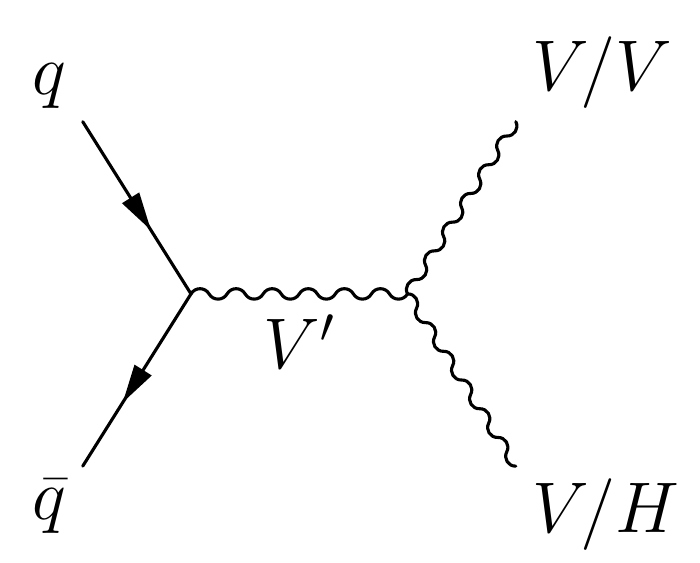}
      	}\\
	\subfloat[]{
      		\includegraphics[width=0.35\textwidth]{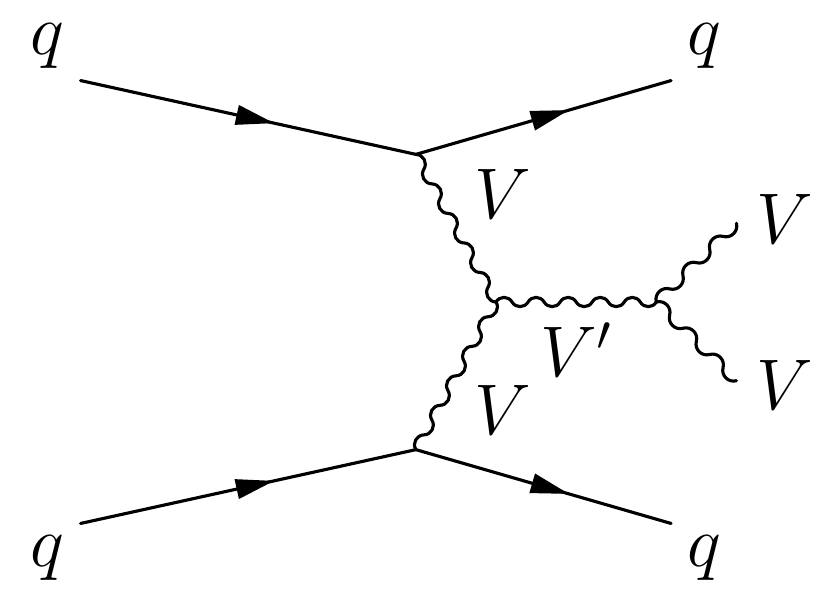}
      	}
        \subfloat[]{
      		\includegraphics[width=0.35\textwidth]{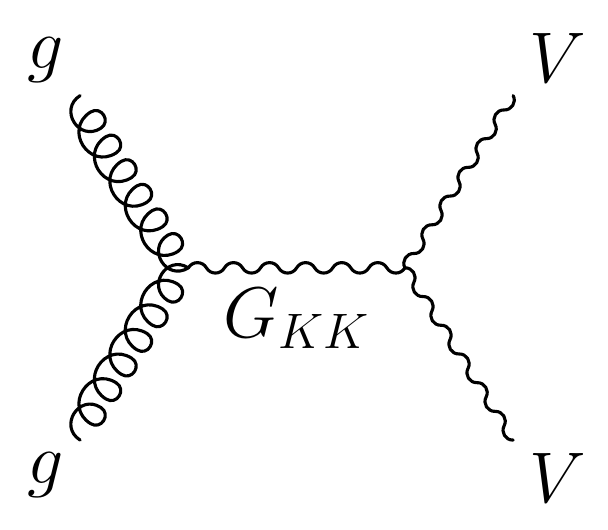}
      	}
  \caption{Feynman diagrams for heavy resonance production and decay: (a) Drell--Yan production and decay into $\ell\nu$/$\ell\ell$, (b) Drell--Yan production and decay into $VV$/$VH$, (c) vector-boson fusion production and decay into $VV$, and (d) gluon--gluon fusion production and decay into $VV$ (with $V = W$ or $Z$).}
\label{fig:FeynmanDiags}
\end{figure}

\section{Signal models}
\label{sec:theory}
The results presented in this article are interpreted in the context of three models: the heavy vector triplet (HVT) model~\cite{HVT,Pappadopulo:2014qza},
the RS model, and an empirical model featuring a new heavy scalar.

The HVT model provides a broad phenomenological framework that encompasses a range of
different scenarios involving new heavy gauge bosons and their couplings to SM fermions and bosons.
In this model, a triplet $\mathcal{W}$ of colorless vector bosons is introduced with zero hypercharge.
This leads to a set of nearly degenerate charged, $W^{\prime \pm}$, and neutral, $Z^{\prime}$, 
states collectively denoted by $V^\prime$.\footnote{The charged state is denoted $W^\prime$ in the remainder of this article.}
For the interpretation performed in this article, the \wp\ and \zp\ masses are taken to be the same.
The model allows one to explore different coupling strengths of those states to quarks,
leptons, vector bosons, and Higgs bosons with the following interaction Lagrangian:
\begin{equation}
\mathcal{L}_\mathcal{W}^\mathrm{int} =  - g_q \mathcal{W}^a_\mu \bar{q}_k\gamma^\mu \frac{\sigma_a}{2} q_k  - g_\ell \mathcal{W}^a_\mu \bar{\ell}_k\gamma^\mu \frac{\sigma_a}{2} \ell_k
- g_H \left(\mathcal{W}^a_\mu H^\dagger \frac{\sigma_a}{2} i D^\mu H   + \mathrm{h.c.}\right),
\label{eq:lagrangian}
\end{equation}
where $q_k$ and $\ell_k$ represent the left-handed quark and lepton doublets for fermion generation $k$
($k = 1, 2, 3$); $H$ represents the Higgs doublet; $\sigma_a$ ($a = 1, 2, 3$) are the Pauli
matrices; and $g_q$, $g_\ell$, and $g_H$ correspond to the coupling strengths between the
triplet field $\mathcal{W}$ and the quark, lepton, and Higgs fields, respectively.\footnote{The 
coupling constants $g_H$, $g_f$, $g_q$, and $g_\ell$ are used in this article. 
They are related to those in Ref.~\cite{Pappadopulo:2014qza} as follows:
the Higgs coupling $g_H = g_V c_H$ and the universal fermion coupling $g_f = g^2 c_F / g_V$, 
where $g$ is the SM SU(2)$_\text{L}$ gauge coupling, while the $c$ parameters and 
the coupling $g_V$ are defined in Ref.~\cite{Pappadopulo:2014qza}.
Couplings specific to quarks and leptons are given by $g_q = g^2 c_q / g_V$ and
$g_\ell = g^2 c_\ell / g_V$.}
Interactions with fermions of different generations are assumed to be universal
and right-handed fermions do not participate.
The triplet field interacts with the Higgs field and 
thus with the longitudinally polarized $W$ and $Z$ bosons 
by virtue of the equivalence theorem~\cite{Cornwall:1974km,Gounaris:1986cr,Chanowitz:1985hj}.
In this framework, the branching fractions for the decays
$W^\prime \to WZ$, $W^\prime \to WH$,
$Z^\prime \to WW$, and $Z^\prime \to ZH$,
are equal for $V^\prime$ masses above 1.5~\TeV\ 
and other neutral diboson final states are either suppressed or forbidden.

Three explicit HVT scenarios are used as benchmarks for interpretation of the results. The first two benchmarks are both Drell-Yan (DY) production mechanisms [Figures~\ref{fig:FeynmanDiags}(a) and \ref{fig:FeynmanDiags}(b)], while the third benchmark proceeds via the vector-boson fusion (VBF) mechanism [Figure~\ref{fig:FeynmanDiags}(c)]. Within the DY processes, two scenarios differently emphasize the relative strengths of $g_H$ and $g_f$. The first DY scenario, referred to as model A, reproduces the phenomenology of weakly coupled models based on an extended gauge symmetry~\cite{Barger:1980ix}. In this case, the couplings are $g_H = -0.56$ and $g_f = -0.55$, with the universal fermion coupling $g_f = g_q = g_\ell$. The second DY scenario, referred to as model B, implements a strongly coupled scenario as in composite Higgs models~\cite{Contino:2011np}  
with $g_H = -2.9$ and $g_f = 0.14$.\footnote{In terms of the coupling constants in the notation of Ref.~\cite{Pappadopulo:2014qza}, the choices for models A and B correspond to $g_V = 1$ and $g_V = 3$, respectively.} In model B, the $V^\prime$ resonances are broader than in the weakly coupled scenario, model A, but remain narrow relative to the experimental resolution. The relative width, $\Gamma / m$, is below 5\% over much of the parameter space explored in this article. Model B is not considered for masses below 1500~\GeV\ because model A is used to extract the acceptance of the combined channels, and the branching fractions to $VV$ and $VH$ differ between models A and B in that mass range. The acceptance for individual channels is the same for models A and B. There is also a second constraint for model B, for masses below 800~\GeV, where it is not compatible with SM precision measurements due to increased mixing between the SM gauge bosons and the heavy vector resonance. 

For the DY process with decay of the $V^\prime$ into lepton--antilepton final states,
branching fractions are largest in model A with
values of approximately 4\% and only about 0.2\% in model B, for each generation taken separately.
In contrast, the branching fractions for decays into individual diboson channels are about 2\% 
in model A, whereas they are close to 50\% in model B.

The third scenario, referred to as model C, is designed to focus solely on the rare process of vector-boson fusion. In this case, the $V'$ resonance couplings are set to $g_H = 1$ and $g_f = 0$. Model C is therefore in a separate phase space domain to models A and B and assumes no DY production.  The interpretation can be extended beyond these three benchmark models by considering 
the two-dimensional parameter space consisting of $g_H$ and $g_f$ (assuming fermion universality) or $g_q$ and $g_\ell$ for a given value of $g_H$. The different production mechanisms and decay modes included here provide sensitivity to different regions of this parameter space, with production via the DY process providing sensitivity to $g_q$ and production via VBF providing sensitivity to $g_H$. Likewise, decays into lepton--antilepton states provide sensitivity to $g_\ell$ whereas decays into diboson states provide sensitivity to $g_H$.

The RS model postulates the existence of a warped extra dimension in which only
gravity propagates as in the original ``RS1'' scenario~\cite{RSmodel} or in which both
gravity and all SM fields propagate as in the ``bulk RS'' scenario~\cite{graviton}.
Propagation in the extra dimension leads to a tower of Kaluza--Klein (KK) excitations of
gravitons (denoted \GKK) and SM fields.
In the bulk RS model considered here, KK gravitons are produced via both quark--antiquark
annihilation and gluon--gluon fusion (ggF), with the latter dominating due to suppressed
couplings to light fermions.
The strength of the coupling depends on \kMPl, where $k$ corresponds to the curvature
of the warped extra dimension and $\Mpl = 2.4 \times 10^{18}$~\GeV{} is the effective
four-dimensional Planck scale.
Both the production cross section and decay width of the KK graviton scale as the square of \kMPl.
For the value $\kMPl = 1$ used in the interpretation, the \GKK resonance width
relative to its mass is approximately 6\%.
The \GKK\ branching fraction is largest for decays into the $\ttbar$ final state, with values
ranging from 42\% for $m(\GKK) = 0.5$~\TeV{} to 65\% for $m(\GKK)$ values above 1~\TeV{}.
Corresponding values for the $WW$ ($ZZ$) final state range from 34\% to 20\% (18\% to 10\%).
Table~\ref{tab:signalCrossSections} presents production cross sections for several heavy resonance
masses in the HVT models A, B, and C, and the bulk RS model.

\begin{table}[t]
\caption{Cross sections for production of heavy resonances of different masses in HVT models A and B
 via the Drell--Yan process, in HVT model C via vector-boson fusion, and in the bulk RS model via gluon--gluon fusion and the Drell--Yan process.}
\begin{center}
\begin{tabular}{l l l l l l l l}
\hline
\multicolumn{1}{c}{ } & \multicolumn{2}{c}{HVT model A} & \multicolumn{2}{c}{HVT model B} & \multicolumn{2}{c}{HVT model C} & Bulk RS \\ 
$m$ & $\sigma (W')$ & $\sigma (Z')$ & $\sigma (W')$ & $\sigma (Z')$ & $\sigma (W')$ & $\sigma (Z')$ & $\sigma (\GKK)$ \\
{[TeV]} & [fb] & [fb]  & [fb] & [fb]  & [fb] & [fb]  & [fb] \\
\hline
1.0       & $2.20 \times 10^4$ & $1.12 \times 10^4$ & $987$              & $510$              & $1.30$ & $0.888$     & $583$ \\
2.6       & $219$              & $100$              & $14.0$             & $6.44$             & $4.78 \times 10^{-3}$ & $3.14 \times 10^{-3}$ & $1.41$ \\
4.0       & $9.49$             & $4.37$             & $0.626$            & $0.288$            & $1.27 \times 10^{-4}$ & $7.92 \times 10^{-5}$ & $3.25 \times 10^{-2}$ \\

\hline
\end{tabular}
\label{tab:signalCrossSections}
\end{center}
\end{table}

The last model considered is an empirical model with a narrow heavy scalar resonance
produced via the ggF and VBF mechanisms and decaying directly into $VV$. The width of this new scalar
is assumed to be negligible compared with the detector resolution, and 
the relative branching fractions for decay into the $WW$ and $ZZ$ final states approximately
follow a $2 : 1$ ratio.
This benchmark is used to explore sensitivity to extended Higgs sectors. 
Table~\ref{tab:decay_signalModels} summarizes the channels considered in
the interpretation for each signal model.

\begin{table}[t]
\caption{Signal models, resonances, and decay modes considered in the combination.}
\begin{center}
\begin{tabular}{l c c c c c c c}
\hline
Model~\textbackslash{}~Decay mode & $WW$  & $WZ$  & $ZZ$  & $WH$  & $ZH$  & $\ell\nu$  & $\ell\ell$ \\
\hline
HVT 		& $Z'$     & $W'$ & 	    & $W'$  & $Z'$  & $W'$	 & $Z'$ \\
Bulk RS  	& \GKK     & 	  & \GKK    & 	    & 	    &		 &      \\
Scalar 		& Scalar   &   	  & Scalar  &	    & 	    &		 &      \\

\hline
\end{tabular}
\label{tab:decay_signalModels}
\end{center}
\end{table}

\section{ATLAS detector}
\label{sec:detector}
\newcommand{\AtlasCoordFootnote}{%
ATLAS uses a right-handed coordinate system with its origin at the nominal interaction point (IP)
in the center of the detector and the $z$-axis along the beam pipe.
The $x$-axis points from the IP to the center of the LHC ring,
and the $y$-axis points upwards.
Cylindrical coordinates $(r,\phi)$ are used in the transverse plane,
$\phi$ being the azimuthal angle around the $z$-axis.
The pseudorapidity is defined in terms of the polar angle $\theta$ as $\eta = -\ln \tan(\theta/2)$.
Angular distance is measured in units of $\Delta R \equiv \sqrt{(\Delta\eta)^{2} + (\Delta\phi)^{2}}$.}

The ATLAS experiment~\cite{PERF-2007-01,ATLAS-TDR-19} at the LHC is a multipurpose particle detector
with a forward--backward symmetric cylindrical geometry and a near $4\pi$ coverage in
solid angle.\footnote{\AtlasCoordFootnote}
It consists of an inner detector for tracking surrounded by a thin superconducting solenoid
providing a \SI{2}{\tesla} axial magnetic field, electromagnetic and hadronic calorimeters, and a muon spectrometer.
The inner detector covers the pseudorapidity range $|\eta| < 2.5$.
It consists of silicon pixel, silicon microstrip, and transition-radiation tracking detectors.
A new innermost pixel layer~\cite{ATLAS-TDR-19} inserted at a radius of 3.3~cm has been used since 2015.
Lead/liquid-argon (LAr) sampling calorimeters provide electromagnetic (EM) energy measurements
with high granularity.
A hadronic (steel/scintillator-tile) calorimeter covers the central pseudorapidity range ($|\eta| < 1.7$).
The endcap and forward regions are instrumented with LAr calorimeters
for both the EM and hadronic energy measurements up to $|\eta| = 4.9$.
The muon spectrometer surrounds the calorimeters and features
three large air-core toroidal superconducting magnet systems with eight coils each.
The field integral of the toroids ranges between \num{2.0} and \SI{6.0}~Tm
across most of the detector. The muon spectrometer includes a system of precision tracking chambers up to $|\eta| = 2.7$ and fast detectors for triggering up to $|\eta| = 2.4$.
A two-level trigger system~\cite{TRIG-2016-01} is used to select events.
The first-level trigger is implemented in hardware and uses a subset of the detector information
to reduce the accepted rate to at most \SI{100}{\kilo\hertz}.
This is followed by a software-based trigger level that reduces the accepted event rate 
to \SI{1}{\kilo\hertz} on average.

\section{Data and Monte Carlo simulation}
\label{sec:MC}
The data sample was collected by the ATLAS detector during the $pp$ collision running of the
LHC at $\rts = 13$~\TeV{} in 2015 and 2016. 
Events were selected for the different channels with various triggers, as described in their 
respective papers~\cite{EXOT-2016-19,EXOT-2016-28,EXOT-2016-11,HIGG-2016-31,EXOT-2016-29,HIGG-2016-19,EXOT-2016-12,EXOT-2016-10,EXOT-2016-06,EXOT-2016-05}. 
Channels featuring charged or neutral leptons were selected with single or multiple electron and muon
triggers with various \pT\ thresholds and isolation requirements, 
or with missing transverse momentum triggers with varying thresholds.
A high-\pt\ jet trigger was used in the fully hadronic channels.
After requiring that the data were collected
during stable beam conditions and with a functional detector, 
the integrated luminosity amounts to 36.1~\ifb.

The interpretation in the combined channels relies on MC
simulation to model the shape and normalization of the signals described in Section~\ref{sec:theory}.
Signal events for the HVT and bulk RS models were generated with \MGMCatNLO v2.2.2~\cite{madgraph} 
at leading order (LO) using the NNPDF23LO parton distribution function (PDF) set~\cite{Ball:2012cx}.
For the production of resonances in the HVT model, both the DY 
and VBF mechanisms were simulated, whereas for 
the bulk RS model, \GKK\ resonances were produced via the ggF and DY mechanisms.
In the case of the heavy scalar model, signal events were generated at next-to-leading order (NLO)
via the ggF and VBF mechanisms with \POWHEGBOX v1~\cite{Alioli:2008tz,Alioli:2010xd}
and the CT10 PDF set~\cite{ct10}.
The ggF/DY and VBF processes were simulated as independent MC samples.
For all signal models and production mechanisms, the generated events were interfaced 
to \PYTHIAV{v8.186}~\cite{pythia8} for parton showering, hadronization, and the underlying event. 
This interface relied on the A14 set of tuned parameters~\cite{ATL-PHYS-PUB-2014-021} 
for events generated with \MGMCatNLO at LO and 
the AZNLO set of tuned parameters~\cite{STDM-2012-23} for events generated with \POWHEGBOX at NLO.
Interference between the signal events and SM processes was not taken into account
as the results for the bosonic channels are expected to change negligibly for the models 
considered since they predict narrow resonances. 
The particular case of $\ell\nu$ and $\ell\ell$ 
channels is discussed in Section~\ref{sec:selection}.
Examples of generator-level signal mass distributions are shown in Figure~\ref{fig:signalMC}.

Simulated background event samples are used to derive the main background estimates in the case of analyses in the 
$\ell\nu$, $\ell\ell$, $\ell\nu\ell\ell$, and $\ell\ell\ell\ell$ channels, and to extrapolate backgrounds from control regions in the analysis of the other channels. 
In other cases, the data are used to extract the normalization and/or shape of the background distributions.
Although the production of background MC samples differed somewhat depending on the specific
analysis, most MC samples were produced as follows.
Diboson ($WW$, $WZ$, $ZZ$) events were generated with \SHERPA~\cite{Gleisberg:2008ta}
or \POWHEGBOX;
$W$+jets and $Z$+jets events were generated with \SHERPA for up to two partons at NLO
and up to four partons at LO using the OpenLoops~\cite{Cascioli:2011va} and Comix~\cite{Gleisberg:2008fv} programs, respectively.
The production of top-quark pairs and single top quarks was performed at NLO with
\POWHEGBOX.
For the $\ell\nu$ and $\ell\ell$ channels, the dominant DY background was modeled using
\POWHEGBOX with next-to-next-to-leading-order QCD and NLO electroweak corrections.
More specific details can be found in the papers for each analysis.

For all MC samples, except those produced with \SHERPA, 
$b$-hadron and $c$-hadron decays were performed with
{\textsc{EvtGen}} v1.2.0~\cite{Lange:2001uf}.
The production of the simulated event samples included the effect of multiple $pp$ interactions 
per bunch crossing, as well as the effect on the detector response due to interactions from bunch 
crossings before or after the one containing the hard interaction. These effects are collectively
referred to as ``pileup.''
The simulation of pileup collisions was performed with \PYTHIA8\ and tuned to reproduce 
the average of 23 pileup interactions observed in the data 
in addition to the hard-scatter interaction.
Most of the MC samples were processed through a detailed simulation of the detector response
with \GEANT~4~\cite{geant, SOFT-2010-01}. A small subset of MC samples was processed 
with a fast parameterization of the calorimeter response~\cite{ATL-PHYS-PUB-2010-013}
while the response for the other detector components used \GEANT~4.
In all cases, events were reconstructed with the same software as was used for the data.

\begin{figure}[tb]
\begin{center}
\includegraphics[width=0.60\linewidth]{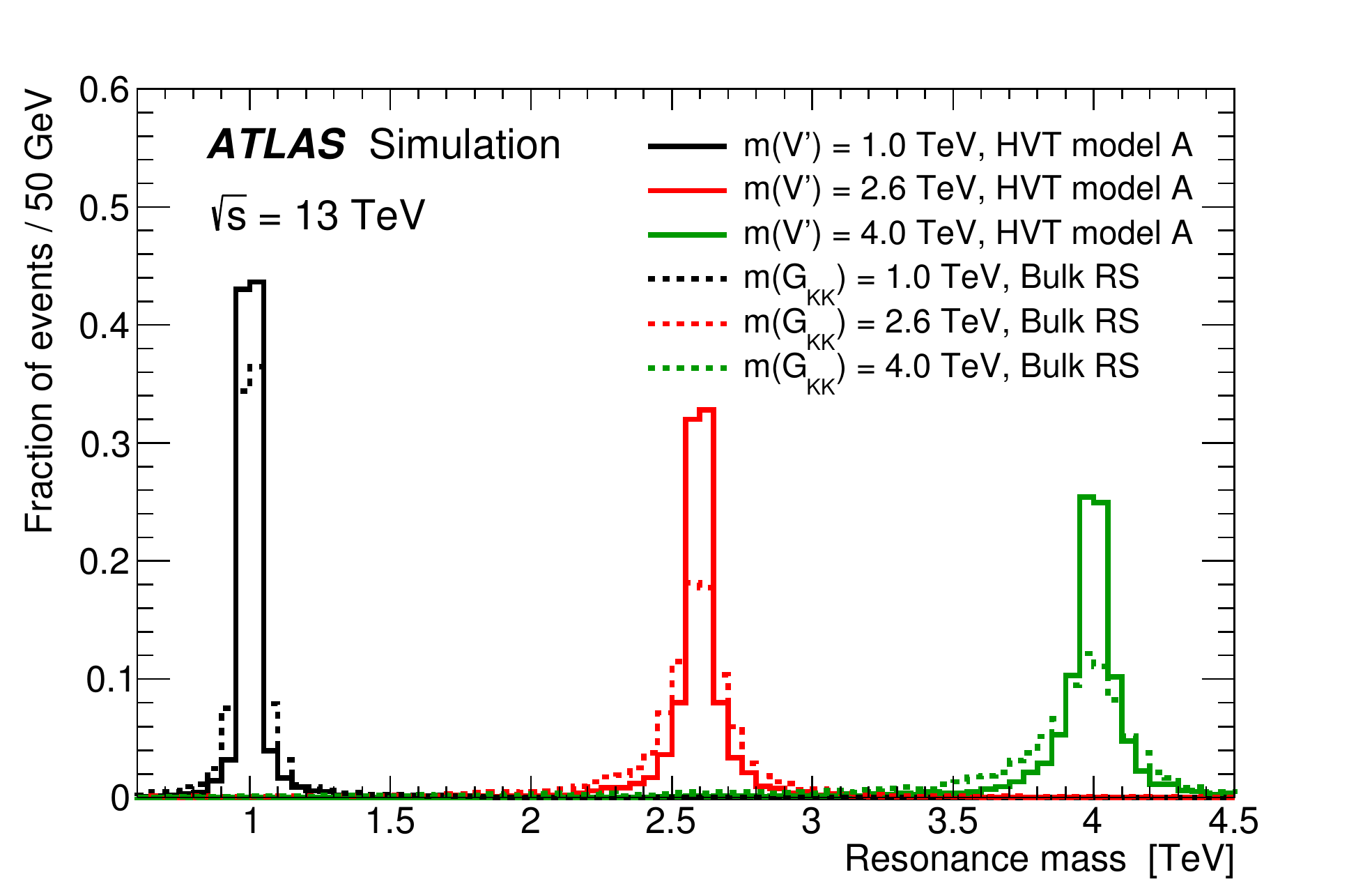}
\end{center}
\vspace{-.4cm}
\caption{Generator-level mass distributions for HVT model A and the bulk RS signal model
 at resonance masses of 1.0, 2.6, and 4.0 \TeV. The histograms are normalized to an area
 equal to unity.}
\label{fig:signalMC}
\end{figure}


\section{Event reconstruction}
\label{sec:reconstruction}
The event selection discussed in Section~\ref{sec:selection} relies on the reconstruction
of electrons, muons, jets, and missing transverse momentum (with magnitude \met). 
Although the requirements vary for the different channels, the general algorithms are
introduced below. The small differences between the efficiencies measured in data and MC simulation 
are corrected for by applying scale factors to the MC simulation so that it matches the data.

Measurements in the inner detector are used to reconstruct tracks from charged particles.
The resulting tracks are then used to reconstruct collision vertices from $pp$ interactions
along the beam axis as well as vertices from the decays of $b$- and $c$-hadrons that are
displaced from that axis. 
Out of the multiple collision vertices in each bunch crossing, a primary vertex is
selected as the vertex with the largest $\sum \pt^2$, where the sum is over all tracks
with transverse momentum $\pt > 0.4$~\GeV{} which are associated with the vertex.
Tracks associated with the primary vertex are identified as electrons or muons if they
satisfy a set of criteria.
Electrons are identified as tracks matching energy clusters in the electromagnetic calorimeter
with energy deposition consistent with that of an electromagnetic shower~\cite{ATLAS-CONF-2016-024}.
In addition, electron candidates must satisfy a set of isolation 
criteria~\cite{ATLAS-CONF-2016-024}. Different tightness levels of identification and isolation
are used depending on the needs of each analysis.
Muons are identified by matching inner detector tracks to full tracks or track segments
reconstructed in the muon spectrometer. Identification and isolation criteria that are
specific to different tightness levels are detailed in Ref.~\cite{PERF-2015-10}.

Jets are reconstructed from clusters of energy deposits in calorimeter cells~\cite{PERF-2014-07}
with the \antikt\ clustering algorithm~\cite{Cacciari:2008gp} implemented in FastJet~\cite{Cacciari:2011ma}. 
To remove jets reconstructed from pileup, jet-vertex tagging (JVT) is applied
to jets with $\pt < 60$~\GeV{} and $|\eta|<2.4$~\cite{PERF-2014-03}.
Jets built using a radius parameter $R$ equal to 0.4 are referred to as
``small-$R$'' jets and those built using $R$ equal to 1.0 are referred to as ``large-$R$'' jets.
A pair of small-$R$ jets may be used to reconstruct $V \to qq$ decays at sufficiently small
$V$ momentum where they can be resolved, but a single large-$R$ jet is used at higher momentum when the two small-$R$ jets 
merge due to the high Lorentz boost.
Small-$R$ jets are built from clusters calibrated at the EM scale~\cite{PERF-2016-04} while
large-$R$ jets are built from clusters calibrated at the local hadronic scale~\cite{PERF-2011-03}. 
The latter jets are trimmed to minimize the impact of pileup and to improve their
energy and mass resolution
by reclustering the constituents of each jet with the \kt\ algorithm~\cite{Ellis:1993tq}
into smaller $R = 0.2$ subjets and removing those subjets with 
$\pt^{\text{subjet}} / \,\pt^{\text{jet}} < 0.05$, where $\pt^{\text{subjet}}$
and $\pt^{\text{jet}}$ are the transverse momenta of the subjet and original
jet, respectively.
Calibration of the trimmed jet \pt\ and mass is described in Ref.~\cite{ATL-PHYS-PUB-2015-033}.

Jets containing $b$-hadron decay products are tagged with a multivariate algorithm that exploits 
the presence of large-impact-parameter tracks and displaced vertices 
from $b$-hadron decays~\cite{PERF-2012-04,ATL-PHYS-PUB-2016-012}.
Large-$R$ jets are tagged as consistent with hadronic decays of $W$ or $Z$ bosons 
based on the mass (the mass window varies with jet p$_T$) and substructure of the jet~\cite{PERF-2015-03,ATL-PHYS-PUB-2015-033}. 
The latter exploits the two-body kinematics of high-\pt\ $V\rightarrow qq$ decays as
measured by the variable $D_2$, which is defined as a ratio of two-point to 
three-point energy correlation functions that are based on the energies of, and 
pairwise angular distances between, the jet's constituents~\cite{Larkoski:2014gra,PERF-2015-03}. 
Likewise, large-$R$ jets may also be tagged as originating from $H \to bb$ decays
by requiring the jet mass to be consistent with that of the Higgs boson (75--145~\GeV) and 
the presence of two or more $R = 0.2$ jets built from tracks associated with the large-$R$ jet,
at least one of which must satisfy $b$-tagging requirements.

The magnitude of the event's missing transverse momentum is computed from the vectorial sum
of the transverse momenta of calibrated electrons, muons, and small-$R$ jets in the event~\cite{PERF-2014-04}. 
The \met\ value is corrected for the soft term, which consists of tracks associated with the primary
vertex but not associated with electrons, muons, taus leptons, or small-$R$ jets.


\section{Event selection}
\label{sec:selection}
The event selection and background estimation for the different analyses are briefly presented
here. A full description is available in Refs.~\cite{EXOT-2016-19,EXOT-2016-28,EXOT-2016-11,HIGG-2016-31,EXOT-2016-29,HIGG-2016-19,EXOT-2016-12,EXOT-2016-10,EXOT-2016-06,EXOT-2016-05}. 
A list of the channels that are input to the combination is provided in Table~\ref{tab:signatures}
along with their experimental signatures. 
Care was taken in defining the event selection to achieve orthogonality between the different channels, as described later in this Section.
The channels are broadly separated into three categories, depending on the targeted decay state of the intermediate resonance: a vector-boson pair ($VV$), a $W$ or $Z$ boson with an associated Higgs boson ($VH$), and a pair of leptons (not involving intermediate bosons). Within the $VV$ category, there are three subcategories: fully hadronic, semileptonic, and fully leptonic. In the semileptonic and fully leptonic subcategories, the searches are further split into optimized selections for ggF/DY and VBF production (only the ggF/DY signature is indicated in Table~\ref{tab:signatures}). The VBF-enriched selections are made orthogonal to the ggF/DY selections by requiring the presence of additional small-$R$ jets, of which the two with the highest \pt\ must have large $\eta$ separation and high invariant mass. For the majority of the searches the discriminating variable is the invariant mass of the $VV$/$VH$/$\ell\ell$ candidates, except those which involve two neutrinos (or the $W'$ $\rightarrow$ $\ell\nu$ final state) where the transverse mass of the final-state particles is used.

\begin{table}[t]
\caption{Summary of analysis channels, diboson states they are sensitive to, and
 their experimental signatures. The selection reflects requirements specific to each channel.
 Additional jets (not included in the ``Jets'' column) are required to define VBF categories.
 The notation ``j'' represents small-$R$ jets and ``J'' represents large-$R$ jets.
 Leptons are either electrons or muons. The notation ``$1e, 1\mu$'' means that the signature is either $1e$ or $1\mu$, whereas ``$1e$+$1\mu$'' means $1e$ and $1\mu$.
 A veto is imposed on \met\ in some channels to guarantee orthogonality between final-state channels.
 The symbol ``$-$'' signifies that no requirement is imposed on a given signature.}
\begin{center}
\begin{tabular}{l l c c c c c c}
\hline
Channel            & Diboson state & \multicolumn{4}{c}{~~~~~~Selection}       & VBF cat. & Reference \\
                   &               & Leptons       & \met  & Jets   & $b$-tags &          &           \\
\hline
$qq qq$            & $WW/WZ/ZZ$    & 0             & veto  & 2J     & $-$      & $-$      & \cite{EXOT-2016-19} \\
$\nu\nu qq$        & $WZ/ZZ$       & 0             & yes   & 1J     & $-$      & yes      & \cite{EXOT-2016-29} \\
$\ell\nu qq$       & $WW/WZ$       & $1e, 1\mu$    & yes   & 2j, 1J & $-$      & yes      & \cite{EXOT-2016-28} \\
$\ell\ell qq$      & $WZ/ZZ$       & $2e, 2\mu$    & $-$   & 2j, 1J & $-$      & yes      & \cite{EXOT-2016-29} \\
$\ell\ell\nu\nu$   & $ZZ$          & $2e, 2\mu$    & yes   & $-$    & 0        & yes      & \cite{HIGG-2016-19} \\
$\ell\nu\ell\nu$   & $WW$          & $1e$+$1\mu$   & yes   & $-$    & 0        & yes      & \cite{HIGG-2016-31} \\
$\ell\nu\ell\ell$  & $WZ$          & $3e$, $2e$+$1\mu$, $1e$+$2\mu$, $3\mu$ & yes & $-$ & 0 & yes & \cite{EXOT-2016-11} \\
$\ell\ell\ell\ell$ & $ZZ$          & $4e$, $2e$+$2\mu$, $4\mu$ & $-$ & $-$ & $-$ & yes    & \cite{HIGG-2016-19} \\
\hline
$qq bb$            & $WH/ZH$       & 0             & veto  & 2J     & 1, 2     & $-$      & \cite{EXOT-2016-12} \\
$\nu\nu bb$        & $ZH$          & 0             & yes   & 2j, 1J & 1, 2     & $-$      & \cite{EXOT-2016-10} \\
$\ell\nu bb$       & $WH$          & $1e, 1\mu$    & yes   & 2j, 1J & 1, 2     & $-$      & \cite{EXOT-2016-10} \\
$\ell\ell bb$      & $ZH$          & $2e, 2\mu$    & veto  & 2j, 1J & 1, 2     & $-$      & \cite{EXOT-2016-10} \\
\hline
$\ell\nu$          & $-$           & $1e, 1\mu$    & yes   & $-$    & $-$      & $-$      & \cite{EXOT-2016-06} \\
$\ell\ell$         & $-$           & $2e, 2\mu$    & $-$   & $-$    & $-$      & $-$      & \cite{EXOT-2016-05} \\
\hline
\end{tabular}
\label{tab:signatures}
\end{center}
\end{table}

Many of the searches involving charged leptons are affected by events with lepton candidates 
that originate from jets misidentified as leptons or with non-prompt leptons that originate
from hadron decays. This background source is referred to as the ``fake-lepton'' background and
is estimated using data-driven techniques. 
Events with fake leptons may arise from a variety of different processes including
multijet, $W/Z$+jets, and $\ttbar$ production.
Other background sources are estimated using MC simulation, with constraints sometimes
extracted from control regions in the data.

The fully hadronic $VV$ final state benefits from the largest branching fraction among the possible final states, but suffers from a large background contamination from the production of multijet events. However, this contamination can be mitigated in the regime of TeV-scale resonances with jet substructure techniques as described in Section~\ref{sec:reconstruction}.
The background prediction is obtained with a fit to the invariant mass distribution of the two
highest-\pt\ large-$R$ jets in the event. This channel explores the mass range between 1.1
and 5.0~\TeV\ and is particularly sensitive at high resonance mass.

The semileptonic $VV$ analyses require either two small-$R$ jets or one large-$R$ jet, for the resolved and merged regimes respectively, in addition to zero, one, or two leptons, with significant \met\ required in all channels except $\ell\ell qq$. Control regions are used to derive the background estimate, and separate signal regions are defined so as to be sensitive to the different production mechanisms, i.e.\ ggF/DY or VBF production. The background in the $\nu \nu qq$ channel has large contributions from $W$/$Z$+jets and $t\bar{t}$ events. The background in the $\ell \nu qq$ channel is dominated by $W$+jets and $t\bar{t}$ events, while the background in the $\ell \ell qq$ channel is dominated by $Z$+jets events. These channels are used in the mass range from 0.3 to 5.0~\TeV\ and
are particularly sensitive in the mid- to high-mass range.

For the fully leptonic $VV$ final states, different selection categories are defined for each channel to optimize the sensitivity to DY, ggF, and VBF production. In the $\ell \nu \ell \nu$ channel, two VBF categories are defined with $N_\text{jet} = 1$ and $N_\text{jet} \ge 2$, with additional criteria on the jet $\eta$ and separation between the leptons and jets to minimize contamination from the ggF signal. A third category for ggF production is further defined as those events that fail to enter the two VBF categories, while satisfying the other base criteria, ensuring orthogonality. The major backgrounds in the $\ell \nu \ell \nu$ channel come from $WW$ and $t\bar{t}$ production. This channel is used in the mass range 0.5--5.0~\TeV\ (0.5--1.0~\TeV) for ggF (VBF) production with particular sensitivity at lower mass. For the $\ell \nu \ell \ell$ channel, two categories are defined to discriminate between DY and VBF production mechanisms. The dominant background in the $\ell \nu \ell \ell$ channel is the contribution from $WZ$ production, and this channel has particular sensitivity in the mass range 0.3--3.0~\TeV\ (0.3--2.0~\TeV) for DY (VBF) production. The $\ell \ell \ell \ell$ channel considers all combinations of electron and muon pairs, with $ZZ$ production as the main background contribution. This channel provides good sensitivity for resonance masses below 1~\TeV\ and covers the range of 0.2--2.0~\TeV. Finally, the $\ell \ell \nu \nu$ channel requires exactly two same-flavor and oppositely charged electrons or muons, with the dilepton invariant mass required to be within the $Z$-mass region. This channel has four signal categories: two for ggF and two for VBF production, divided according to the flavor of the leptons they contain. This channel covers the resonance mass range of 0.3--2.0~\TeV, with a particular sensitivity at low mass, between 0.5 and 1.0~\TeV.

The fully hadronic $VH$ analysis focuses on resonance masses above 1~\TeV, with highly boosted $V$ bosons and Higgs bosons that are likely to be highly collimated and merged into a single large-$R$ jet. The analysis uses dedicated boosted-boson tagging and only considers the merged regime, requiring at least two large-$R$ jets with high \pt, with a veto on any event that contains a lepton candidate. The main background in this search comes from multijet processes.

The semileptonic $VH$ analyses focus on the resonance mass region above 0.5~\TeV. 
Regimes in which the $V$ or $H$ boson decay constituents are separated enough to be considered resolved and those in which they are merged are both considered in separate categories, with priority given to the resolved analysis, and the remaining events recycled into the merged analysis. The semileptonic searches are split into three channels depending on the number of charged leptons: $\nu \nu bb$, $\ell \nu bb$, and $\ell \ell bb$.

The $\ell\nu$ and $\ell\ell$ final states have a high sensitivity due to their very clean signature, with good lepton energy resolution and relatively low background. The dominant background in these channels comes from the irreducible charged-current (CC) and neutral-current (NC) DY processes for the $\ell \nu$ and $\ell \ell$ channels, respectively. These searches are sensitive across a wide range of resonance masses from 0.2 to 5.5~\TeV.

For a number of the signatures there is interference between the signal and the SM background. For some channels such as the hadronic and semileptonic diboson decay channels, the impact of interference is expected to be negligible because the SM diboson background is small. Moreover, multijet event production is depleted in $qqqq$ states, and thus the interference with the fully hadronic decay channel is reduced. For the fully leptonic diboson decay channels, this background is not negligible, but the role of interference, which increases with the heavy resonance width, is small for widths less than 15\% of the resonance pole mass. Since only narrow resonances are considered, the impact of interference is neglected. Finally, for the leptonic channels ($\ell\nu$, $\ell\ell$) the interference can play an important role as the dominant background is the irreducible DY process which interferes with the HVT signal, and thus to minimize the effects of interference the $\ell\nu$ transverse mass is required to satisfy $|m_\text{T} - m_\text{pole}| < \sqrt{64~\GeV \times m_\text{pole}}$ in the $\ell\nu$ channel, where $m_\text{pole}$ corresponds to the \wp\ pole mass. Likewise, the $\ell\ell$ mass is required to satisfy $|m_{\ell\ell} - m_\text{pole}| < \sqrt{25~\GeV \times m_\text{pole}}$ in the $\ell\ell$ channel with $m_\text{pole}$ the \zp\ pole mass. The mass window requirement results in the difference between the theoretical cross section with and without interference being less than 15\% throughout the coupling plane.

Possible overlaps among the different searches in the combination are considered to ensure orthogonality. The first step is to determine the orthogonality of the selection criteria used in the various analyses, which is summarized in Table~\ref{tab:signatures}. One of the criteria that cleanly provides orthogonality is the requirement on the lepton multiplicity in the selected events, ranging from zero to four leptons. Further orthogonality is achieved with additional selection criteria for the jets and \met\ in the events. In particular, a veto is applied to events with a large \met\ value in the $qqqq$, $qqbb$, and $\ell\ell bb$ channels. For the combination of $VV$ and $VH$ channels (and also with the leptonic channels), events are further removed from the $VH$ analysis if they are in overlapping parts of the signal region and have a Higgs boson candidate mass close to the $W$/$Z$ mass. This has the effect of improving the $VH$ sensitivity in the combination above 1~\TeV\ by 10--15\% because the original $VH$ semileptonic analyses were optimized for resonances with a mass below 1~\TeV. Only a negligible number of events that overlap between channels remains.

\section{Systematic uncertainties}
\label{sec:systematics}
The various sources of experimental and theoretical systematic uncertainty are assessed as a function of the discriminating variable in each of the search channels in the combination. These uncertainties are derived for both the signal and background estimates where relevant and are treated as correlated or uncorrelated between the signal and background in the various channels, as appropriate. The systematic uncertainties estimated to have a non-negligible impact on the expected cross-section limit are used as nuisance parameters in the statistical interpretation, as described in Section~\ref{sec:statistical}. This section describes the systematic uncertainties for all channels in the combination, and applies to the various signal scenarios in Table~\ref{tab:decay_signalModels}. 
A full description of the evaluation of systematic uncertainties is provided in the original publications for each of the analyses. What follows is a qualitative discussion.

The experimental systematic uncertainties related to charged leptons, such as the efficiencies due to triggering, reconstruction, identification, and isolation, as well as the lepton energy scale and resolution, are evaluated using $Z \rightarrow \ell\ell$ decays and then extrapolated to higher energies. These uncertainties are correlated between leptons of the same flavor across all channels in the combination, and between the signal and the background estimates. The systematic uncertainties for each of the channels featuring charged leptons are summarized in Table~\ref{tab:lep_sys} including the assumed correlation between channels.

The experimental systematic uncertainties due to the missing transverse momentum are summarized in Table~\ref{tab:met_sys}. These relate to the \met\ trigger as well as the \met\ scale and resolution, which are estimated in control regions using the data.

The small-$R$ jet uncertainties are relevant for most of the channels in the combination, including those with leptonic final states that contain at least one neutrino, due to the impact of those uncertainties on the \met\ measurement. The uncertainties in the jet energy scale and resolution are derived by comparing the response between the data and the simulation in various kinematic regions and event topologies. Additional contributions to this uncertainty come from the dependence on the pileup activity and on the flavor composition of the jets as well as the punch-through of the energy from the calorimeter into the muon spectrometer. An uncertainty in the efficiency for jets to satisfy the JVT requirements is assessed. The small-$R$ jet uncertainties are summarized in Table~\ref{tab:jet_sys_1}. For large-$R$ jets, the uncertainties in the energy, mass, and $D_2$ scales are estimated by comparing the ratio of calorimeter-based to track-based jet \pt\ measurements in dijet events between the data and the simulation. The uncertainties in the jet mass resolution and jet energy resolution as well as $D_2$ are assessed by applying additional smearing of the jet observables according to the uncertainty in their resolution measurements. A summary of the large-$R$ jet systematic uncertainties is provided in Table~\ref{tab:fatjet_sys_1}.

The flavor-tagging uncertainty is evaluated by varying the data-to-MC corrections in various kinematic regions, based on the measured tagging efficiency and mistag rates. These variations are applied separately to $b$-hadron jets, $c$-hadron jets, and light (quark or gluon) jets, leading to three uncorrelated systematic uncertainties. An additional uncertainty is included due to the extrapolation for the jets with \pt\ beyond the kinematic reach of the data calibration. The flavor-tagging uncertainties are summarized in Table~\ref{tab:ft_sys_1}.

The theoretical uncertainties are split among the various backgrounds, which play greater or lesser roles in each of the search channels, depending on the composition of backgrounds in a given channel. The dominant background in the $\ell\nu$ and $\ell\ell$ channels is from the CC and NC DY processes, respectively. In these channels, theoretical uncertainties arise from PDFs and electroweak corrections. The PDF uncertainties are divided into PDF eigenvector variations, the choice of the nominal PDF set (CT14NNLO~\cite{Dulat:2015mca}) from a number of different PDF sets, as well as the choice of PDF renormalization and factorization scales, and $\alpha_\text{S}$. In the case of the $\ell\ell$ channel, an additional uncertainty due to photon-induced corrections to the NC DY process is also assessed. Similar sources of uncertainty are assessed and included where relevant for other backgrounds such as: top-quark, diboson, $V$+jets, as well as for the multijet background, when an MC-based estimation is used. Specifically, when ``cross section'' uncertainties are mentioned for these backgrounds they refer to cross-section calculations, while ``modeling'' refers to event generator and parton shower comparisons, and ``extrapolation'' refers to the background being extrapolated from a control region to a higher-mass region. One exception is the multijet-modeling systematic uncertainty for channels that include leptons, such as $\ell\nu$ and $\ell\ell$. In these cases the systematic uncertainty includes variations of the data-driven methodology used to derive the fake-lepton background estimate and its subsequent extrapolation to higher masses. All uncertainties are summarized in Table~\ref{tab:theory_sys}. Theoretical uncertainties that affect the acceptance of the signal are also assessed, such as initial- and final-state radiation, PDF variation, and PDF choice. These generally have a negligible impact on the result but are included where relevant in the statistical interpretation.

All channels include an uncertainty in the integrated luminosity of 3.2\% derived following a methodology similar to that detailed in Ref.~\cite{DAPR-2013-01}. This uncertainty is taken to be correlated across the channels and between the signal and background. The uncertainty due to pileup is also considered when it does not have a negligible impact on the analysis, to cover the difference between the ratios of predicted and measured inelastic cross-section values.

For most of the $VV$ and $VH$ analyses, MC-modeling systematic uncertainties play the dominant role in the theoretical uncertainty, while for the leptonic channels, the PDF variation and PDF choice are by far the most dominant. For the experimental systematic uncertainties, analyses selecting jets are most sensitive to systematic uncertainties in the modeling of large-$R$ jets, while the leptonic channels are affected mostly by the uncertainty in the muon reconstruction efficiency and electron isolation efficiency.

\begin{table}
\caption{Lepton systematic uncertainties. The abbreviations ``S'' and ``B'' stand for signal and background, respectively, and ``negl.'' denotes uncertainties that are negligible. Each uncertainty is considered as correlated between the channels listed.}
\small
\begin{center}
\begin{tabular}{lcccccccccc}
\hline
Source 		& $\ell\nu qq$ 	& $\ell\ell qq$ 	& $\ell\ell\nu\nu$ 	& $\ell\nu\ell\nu$ 	& $\ell\nu\ell\ell$ 	& $\ell\ell\ell\ell$ 	& $\ell\nu bb$ 	& $\ell\ell bb$ 	& $\ell\nu$ 	& $\ell\ell$ 	 \\
\hline
Electron trigger 	& S+B 	& S+B 	& S+B 	& S+B 	&    negl.    	& S+B 	& S+B 	& S+B 	&    negl.    	&    negl.    \\ 
Electron reconstruction 	& S+B 	& S+B 	& S+B 	& S+B 	& S+B 	& S+B 	& S+B 	& S+B 	&    negl.    	&    negl.    \\ 
Electron identification 	& S+B 	& S+B 	& S+B 	& S+B 	& S+B 	& S+B 	& S+B 	& S+B 	&    negl.    	& S+B \\ 
Electron isolation 	& S+B 	& S+B 	& S+B 	& S+B 	& S+B 	& S+B 	& S+B 	& S+B 	&    negl.    	& S+B \\ 
Electron energy scale 	& S+B 	& S+B 	& S+B 	& S+B 	& S+B 	& S+B 	& S+B 	& S+B 	& S+B 	& S+B \\ 
Electron energy resolution 	& S+B 	& S+B 	& S+B 	& S+B 	& S+B 	& S+B 	& S+B 	& S+B 	&    negl.    	& S+B \\ 
Muon trigger 		& S+B 	& S+B 	& S+B 	& S+B 	&    negl.    	& S+B 	& S+B 	& S+B 	& S+B 	&    negl.    \\ 
Muon reconstruction 	& S+B 	& S+B 	& S+B 	& S+B 	& S+B 	& S+B 	& S+B 	& S+B 	& S+B 	& S+B \\ 
Muon isolation 		& S+B 	& S+B 	& S+B 	& S+B 	& S+B 	& S+B 	& S+B 	& S+B 	&    negl.    	& S+B \\ 
Muon momentum scale 	& S+B 	& S+B 	& S+B 	& S+B 	& S+B 	& S+B 	& S+B 	& S+B 	&    negl.    	&    negl.    \\ 
Muon momentum resolution 	& S+B 	& S+B 	& S+B 	& S+B 	& S+B 	& S+B 	& S+B 	& S+B 	& S+B 	& S+B \\ 
\hline
\end{tabular}
\label{tab:lep_sys}
\end{center}
\end{table}

\begin{table}
\caption{\met\ systematic uncertainties. The abbreviations ``S'' and ``B'' stand for signal and background, respectively, while ``--'' denotes uncertainties that are not applicable. Each uncertainty is considered as correlated between the channels listed.}
\begin{center}
\begin{tabular}{lcccccccccc}
\hline
Source 		& $\nu\nu qq$ 	& $\ell\nu qq$ 	& $\ell\ell\nu\nu$ 	& $\ell\nu\ell\nu$ 	& $\ell\nu\ell\ell$ 	& $\nu\nu bb$ 	& $\ell\nu bb$ 	& $\ell\nu$ 	 \\
\hline
\met\ trigger 		& S+B 	& S+B 	& S+B 	&    --    	&    --    	& S+B 	& S+B 	&    --    \\ 
\met\ soft-term scale 	& S+B 	& S+B 	& S+B 	& S+B 	& S+B 	& S+B 	& S+B 	& S+B \\ 
\met\ soft-term resolution 	& S+B 	& S+B 	& S+B 	& S+B 	& S+B 	& S+B 	& S+B 	& S+B \\ 
\hline
\end{tabular}
\label{tab:met_sys}
\end{center}
\end{table}

\begin{table}
\caption{Small-$R$ jet systematic uncertainties. The abbreviations ``S'' and ``B'' stand for signal and background, respectively, and ``negl.'' denotes uncertainties that are negligible. Each uncertainty is considered as correlated between the channels listed.}
\begin{center}
\begin{tabular}{lcccccccccc}
\hline
Source 		& $\nu\nu qq$ 	& $\ell\nu qq$ 	& $\ell\ell qq$ 	& $\ell\ell\nu\nu$ 	& $\ell\nu\ell\nu$ 	& $\ell\nu\ell\ell$ 	& $qqbb$ 	& $\nu\nu bb$ 	& $\ell\nu bb$ 	& $\ell\nu$ 	 \\
\hline
Small-$R$ jet energy scale 	& S+B 	& S+B 	& S+B 	& S+B 	& S+B 	& S+B 	& S+B 	& S+B 	& S+B 	& S+B \\ 
Small-$R$ jet energy resolution 	& S+B 	& S+B 	& S+B 	& S+B 	& S+B 	& S+B 	& S+B 	& S+B 	& S+B 	& S+B \\ 
Small-$R$ jet flavor 	& S+B 	& S+B 	& S+B 	& S+B 	& S+B 	&    negl.    	& S+B 	& S+B 	& S+B 	& S+B \\ 
Small-$R$ jet pileup 	& S+B 	& S+B 	& S+B 	& S+B 	& S+B 	&    negl.    	& S+B 	& S+B 	& S+B 	& S+B \\ 
Small-$R$ jet punch-through 	& S+B 	& S+B 	& S+B 	& S+B 	& S+B 	&    negl.    	& S+B 	& S+B 	& S+B 	& S+B \\ 
Small-$R$ jet JVT 	& S+B 	& S+B 	& S+B 	& S+B 	& S+B 	& S+B 	& S+B 	& S+B 	& S+B 	& S+B \\ 
\hline
\end{tabular}
\label{tab:jet_sys_1}
\end{center}

\end{table}

\begin{table}
\caption{Large-$R$ jet systematic uncertainties. The abbreviations ``S'' and ``B'' stand for signal and background, respectively. Each uncertainty is considered as correlated between the channels listed.}
\small
\begin{center}
\begin{tabular}{lcccccccc}
\hline
Source 		& $qqqq$ 	& $\nu\nu qq$ 	& $\ell\nu qq$ 	& $\ell\ell qq$ 	& $qq bb$ 	& $\nu\nu bb$ 	& $\ell\nu bb$ 	& $\ell\ell bb$ \\
\hline
Large-$R$ jet $D_2$ scale 	& S	  	& S+B 	& S+B 	& S+B 	& S+B 	& S+B 	& S+B 	& S+B \\ 
Large-$R$ jet $D_2$ resolution 	& S	  	& S+B 	& S+B 	& S+B 	& S+B 	& S+B 	& S+B 	& S+B \\ 
Large-$R$ jet scale 	& S	  	& S+B 	& S+B 	& S+B 	& S+B 	& S+B 	& S+B 	& S+B \\ 
Large-$R$ jet resolution 	& S	  	& S+B 	& S+B 	& S+B 	& S+B 	& S+B 	& S+B 	& S+B \\ 
Large-$R$ jet mass scale 	& S	  	& S+B 	& S+B 	& S+B 	& S+B 	& S+B 	& S+B 	& S+B \\ 
Large-$R$ jet mass resolution 	& S	  	& S+B 	& S+B 	& S+B 	& S+B 	& S+B 	& S+B 	& S+B \\ 
\hline
\end{tabular}
\label{tab:fatjet_sys_1}
\end{center}

\end{table}

\begin{table}
\caption{Flavor-tagging systematic uncertainties. The abbreviations ``S'' and ``B'' stand for signal and background, respectively. Each uncertainty is considered as correlated between the channels listed.}
\begin{center}
\begin{tabular}{lcccccccccc}
\hline
Source 		& $\nu\nu qq$ 	& $\ell\nu qq$ 	& $\ell\ell qq$ 	& $\ell\ell\nu\nu$ 	& $\ell\nu\ell\nu$ 	& $\ell\nu\ell\ell$ 	& $qqbb$ 	& $\nu\nu bb$ 	& $\ell\nu bb$ 	& $\ell\ell bb$ \\
\hline
$b$-tagging 		& S+B 	& S+B 	& S+B 	& B	  	& B	  	& B	  	& S+B 	& S+B 	& S+B 	& S+B \\ 
$c$-tagging 		& S+B 	& S+B 	& S+B 	& B	  	& B	  	& B	  	& S+B 	& S+B 	& S+B 	& S+B \\ 
Light-$q$ tagging 	& S+B 	& S+B 	& S+B 	& B	  	& B	  	& B	  	& S+B 	& S+B 	& S+B 	& S+B \\ 
Tagging extrapolation 	& S+B 	& S+B 	& S+B 	& B	  	& B	  	& B	  	& S+B 	& S+B 	& S+B 	& S+B \\ 
\hline
\end{tabular}
\label{tab:ft_sys_1}
\end{center}

\end{table}

\begin{table}
\caption{Theoretical systematic uncertainties. The abbreviation ``B'' stands for background, while ``--'' denotes uncertainties that are not applicable, ``negl.'' denotes uncertainties that are negligible, and ``Corr'' marks whether the uncertainty is correlated between the channels listed. The abbreviation ``F'' means that this parameter was left to float in the background control region for that channel. The systematic uncertainties in the background modeling for the fully hadronic analysis $qqqq$ are embedded in the fit function used to model the background.}
\small
\begin{center}
\begin{tabular}{lccccccccccccc}
\hline
Source 		& Corr 	& $\nu\nu qq$ 	& $\ell\nu qq$ 	& $\ell\ell qq$ 	& $\ell\ell\nu\nu$ 	& $\ell\nu\ell\nu$ 	& $\ell\nu\ell\ell$ 	& $qqbb$ 	& $\nu\nu bb$ 	& $\ell\nu bb$ 	& $\ell\ell bb$ 	& $\ell\nu$ 	& $\ell\ell$ 	 \\
\hline
DY PDF variation 	& Yes 	&    --    	&    --    	&    --    	&    --    	&    --    	&    --    	&    --    	&    --    	&    --    	&    --    	& B 	& B \\ 
DY PDF choice 		& Yes 	&    --    	&    --    	&    --    	&    --    	&    --    	&    --    	&    --    	&    --    	&    --    	&    --    	& B 	& B \\ 
DY PDF scale 		& Yes 	&    --    	&    --    	&    --    	&    --    	&    --    	&    --    	&    --    	&    --    	&    --    	&    --    	&    negl.    	& B \\ 
DY $\alpha_\text{S}$ 		& Yes 	&    --    	&    --    	&    --    	&    --    	&    --    	&    --    	&    --    	&    --    	&    --    	&    --    	& B 	& B \\ 
DY EW corrections 	& Yes 	&    --    	&    --    	&    --    	&    --    	&    --    	&    --    	&    --    	&    --    	&    --    	&    --    	& B 	& B \\ 
DY photon-induced 	& Yes 	&    --    	&    --    	&    --    	&    --    	&    --    	&    --    	&    --    	&    --    	&    --    	&    --    	&    --    	& B \\ 
Top cross section 	& No 	& B 	&    F    	& F 	& B 	& B 	&    --    	& B 	& B 	& B 	& B 	& B 	&    negl.    \\ 
Top extrapolation 	& No 	&    --    	&    --    	&    --    	&    --    	&    --    	&    --    	&    --    	&    --    	&    --    	&    --    	& B 	&    --    \\ 
Top modeling 		& No 	& B 	& B 	& B 	& B 	& B 	&    --    	&    --    	& B 	& B 	& B 	&    negl.    	&    negl.    \\ 
Diboson cross section 	& No 	& B 	& B 	& B 	& B 	& B 	&    --    	& B 	& B 	& B 	& B 	&    negl.    	&    negl.    \\ 
Diboson extrapolation 	& No 	&    --    	&    --    	&    --    	&    --    	&    --    	&    --    	&    --    	&    --    	&    --    	&    --    	& B 	&    --    \\ 
Multijet cross section 	& No 	&    --    	& B 	&    --    	&    --    	&    --    	& B 	& B 	&    --    	& B 	&    --    	&    --    	&    --    \\ 
Multijet modeling 	& No 	&    --    	&    --    	&    --    	&    --    	&    --    	&    --    	& B 	&    --    	& B 	&    --    	& B 	& B \\ 
$Z$+jets cross section 	& No 	&    F    	& B 	&    F    	&    --    	&    --    	&    --    	&    --    	& B 	& B 	& B 	&    --    	&    --    \\ 
$Z$+jets modeling 	& No 	& B 	& B 	& B 	&    --    	&    --    	&    --    	&    --    	& B 	& B 	& B 	&    --    	&    --    \\ 
$W$+jets cross section 	& No 	& B 	&    F    	& B 	&    --    	&    --    	&    --    	&    --    	& B 	& B 	& B 	&    --    	&    --    \\ 
$W$+jets modeling 	& No 	& B 	& B 	& B 	&    --    	&    --    	&    --    	&    --    	& B 	& B 	& B 	&    --    	&    --    \\ 
\hline
\end{tabular}
\label{tab:theory_sys}
\end{center}
\end{table}

\clearpage

\section{Statistical treatment}
\label{sec:statistical}
The combination of the individual channels proceeds with a simultaneous analysis of the signal discriminants across all of the channels. For each signal model being tested, only the channels sensitive to that hypothesis are included in the combination. The statistical treatment of the data is based on the RooFit~\cite{Verkerke:2003ir}, RooStats~\cite{Moneta:2010pm}, and HistFactory~\cite{Cranmer:2012sba} data modeling and handling toolkits.  Results are calculated in two different signal parameterization paradigms, corresponding to one-dimensional upper limits on the cross section times branching fraction (\sigbr) and two-dimensional limits on coupling strengths.  The statistical treatment of each case is described here.

\subsection{One-dimensional upper limits}
 
In the case of one-dimensional upper limits on \sigbr, the overall signal strength, $\mu$, defined as a scale factor multiplying the cross section times branching fraction predicted by the signal hypothesis, is the parameter of interest. The analysis follows the frequentist approach with a test statistic based on the profile-likelihood ratio~\cite{Cowan:2010js}. This test statistic ($\mathcal{T}$) is defined as twice the negative logarithm of the ratio of the conditional (fixed-$\mu$) maximum likelihood to the unconditional maximum likelihood, each obtained from a fit to the data:

\begin{equation}
\label{eqn:LLRnom}
\mathcal{T} = -2 \ln { L(\mu,\hat{\hat{\bm{\theta}}}(\mu)) \over L(\hat{\mu},\hat{\bm{\theta}}(\hat{\mu}))},
\end{equation}

where ${\boldmath \theta}(\mu)$ represent the nuisance parameters. The latter are represented in the equation as their unconditional and conditional maximum-likelihood values, $\hat{\bm{\theta}}(\hat{\mu})$ and $\hat{\hat{\bm{\theta}}}(\mu)$.  The fitted signal strength, $\hat{\mu}$, is bounded from below at zero.

The likelihood, ${L}$, is given by
 
\begin{equation*}
{L} = \prod_{c} \prod_{i} \mathrm{Pois}\left(n^{\mathrm{obs}}_{ci} \left| \, n^\mathrm{sig}_{ci}(\mu, \vec{\theta}) + n^{\mathrm{bkg}}_{ci}( \vec{\theta}) \right. \right) \prod_{k} f_{k}(\theta_k)~,
\end{equation*}

where the index $c$ represents the analysis channel, $i$ represents the bin in the signal discriminant distribution, $n^{\mathrm{obs}}$ is the observed number of events, $n^\mathrm{sig}$ is the number of expected signal events, $n^\mathrm{bkg}$ is the expected number of background events, $\vec{\theta}$ is the vector of nuisance parameters, and $\mathrm{Pois}(x|y)$ is the Poisson probability to observe $x$ events when $y$ are predicted. 

The effect of a systematic uncertainty $k$ on the binned likelihood is modeled with an associated nuisance parameter, $\theta_{k}$, constrained with a corresponding probability density function $f_k(\theta_k)$. In this manner, correlated effects across the different channels are modeled by the use of a common nuisance parameter and its corresponding probability density function. The $f_k(\theta_k)$ terms are Poisson distributed for bin-by-bin MC statistical uncertainties and Gaussian distributed for all other terms.

Given the large number of search channels included in the likelihood, the sampling distribution of the profile-likelihood test statistic is assumed to follow the chi-squared ($\chi^2$) distribution, and thus asymptotic formulae for the evolution of the likelihood as a function of signal strength ($\mu$) are used~\cite{Cowan:2010js}.  In certain instances, such as high-mass tails of resonant mass distributions, the asymptotic approximation is expected to be less reliable.  In these cases, MC trials are used to assess its accuracy. This approximation is found to lead to \sigbr\ limits that are stronger than those obtained with MC trials. The effect is largest in the case of the lepton--antilepton combination for which it increases linearly with resonance mass from approximately 20\% at 2~\TeV\ to 55\% at 5~\TeV.
In the context of HVT model A, the impact of using the asymptotic approximation in the limit setting is at most 250 GeV on the mass limits, as obtained for the lepton-antilepton combination.

When evaluating limits in the HVT model with degenerate-mass \wp\ and \zp\ production, each of the contributing signal processes is normalized to the \sigbr\ value predicted by HVT model A, thereby defining the relative ratios of $\sigma(pp \rightarrow W^\prime)/\sigma(pp \rightarrow Z^\prime)$, $\mathcal{B}(W^\prime \rightarrow \ell \nu)/\mathcal{B}(W^\prime \rightarrow WZ/WH)$ and $\mathcal{B}(Z^\prime \rightarrow \ell \ell)/\mathcal{B}(Z^\prime \rightarrow WW/ZH)$. The HVT model A benchmark makes a model-dependent assumption about these relative ratios, so the resulting upper limits cannot be directly interpreted as general limits on \sigbr.  Thus, the upper limits are presented as the ratio to the HVT model A prediction for \sigbr, as each of the previously defined ratios is fixed to this benchmark prediction.

Upper limits on $\mu$ for the signal models being tested at the simulated resonance masses are evaluated at the 95\% confidence level (CL) following the CL$_\text{s}$ prescription~\cite{Read:2002hq}. Lower limits on the mass of new resonances in these models are obtained by finding the maximum resonance mass where the 95\% CL upper limit on $\mu$ is less than or equal to unity. This mass is found by interpolating  between the limits on $\mu$ at the simulated signal masses. The interpolation assumes monotonic and smooth behavior of the efficiencies for the signal and background processes, and that the impact of the variation of signal mass distributions between adjacent test masses is negligible.

\subsection{Two-dimensional limits}

When calculating one-dimensional upper limits on \sigbr, each of the signal rate predictions from \wp\ and \zp\ production is fixed to the ratio predicted by the benchmark models. To evaluate two-dimensional constraints on coupling strengths, the signal yields are parameterized with a set of coupling parameters ($\vec{g}$) which allow the relative proportions of each signal to vary independently.   Thus, in the two-dimensional limit calculation, Eq.~(\ref{eqn:LLRnom}) is modified to allow the set of coupling parameters to be considered independently:

\begin{equation*}
\mathcal{T}^\prime = -2 \ln { L(\vec{g},\hat{\hat{\bm{\theta}}}(\vec{g})) \over L(\hat{\vec{g}},\hat{\bm{\theta}}(\hat{\vec{g}}))}.
\end{equation*}

The coupling parameterization assumes that all signal production proceeds via quark--antiquark annihilation (proportional to $g_q^2$) and the signal decays are proportional to the square of the bosonic coupling ($g_H$) and leptonic coupling ($g_{\ell }$) in  the $V^\prime \rightarrow VV/VH$  and $V^\prime \rightarrow \ell\nu / \ell\ell$ final states, respectively.  

Two coupling spaces are considered.  The first coupling scenario makes the assumption of common fermionic couplings ($g_f = g_\ell = g_q$) and probes the $\{g_H, g_f\}$ plane.   The second coupling scenario allows independent fermionic couplings and probes the $\{g_q, g_\ell\}$ plane with either  $g_H=0$ or $g_H=-0.56$, where the latter takes the value predicted in the HVT model A benchmark. The 95\% CL limit contours in each coupling space are determined using $\mathcal{T}^\prime$ by normalizing signal rates to the \sigbr\ predictions of the HVT model for the specified values of $\vec{g}$ at a given point in the space and calculating the value of CL$_\text{s}$ for that point.  Upper limits on coupling parameters are thus defined by contours of constant CL$_\text{s}$ in each coupling space considered.


\section{Results}
\label{sec:results}
The methodology described in the previous section is used to statistically combine various channels for the different signal models listed in Table~\ref{tab:decay_signalModels}. The largest local excess is observed in the VBF Scalar ($WW$+$ZZ$) search for a mass of 1.6~\TeV, with a significance of 2.9~$\sigma$. Limits are set on the signal parameters of interest.

For the $VV$ combination, the HVT, bulk RS, and scalar models are all considered. Figure~\ref{fig:hvtLims_WWorWZ} shows the \sigbr\ limit relative to the predicted \sigbr\ for the combination of $W' \rightarrow WZ$ and $Z' \rightarrow WW$ decays in the context of the HVT model for either DY or VBF production mechanisms. Cross-section limits obtained exclusively for the VBF production mechanism are useful for constraining models with small coupling between fermions and $V^\prime$ resonances. Figures~\ref{fig:rsgLims_WWorZZ} and \ref{fig:scalarLims_WWorZZ} show the \sigbr\ limits for the combination of $\GKK \rightarrow WW \;\textrm{or}\; ZZ $ and $\text{Scalar} \rightarrow WW \; \textrm{or}\; ZZ$, respectively.

\begin{figure}[tb]
  \begin{center}
    \subfloat[]{
      \includegraphics[width=0.485\textwidth]{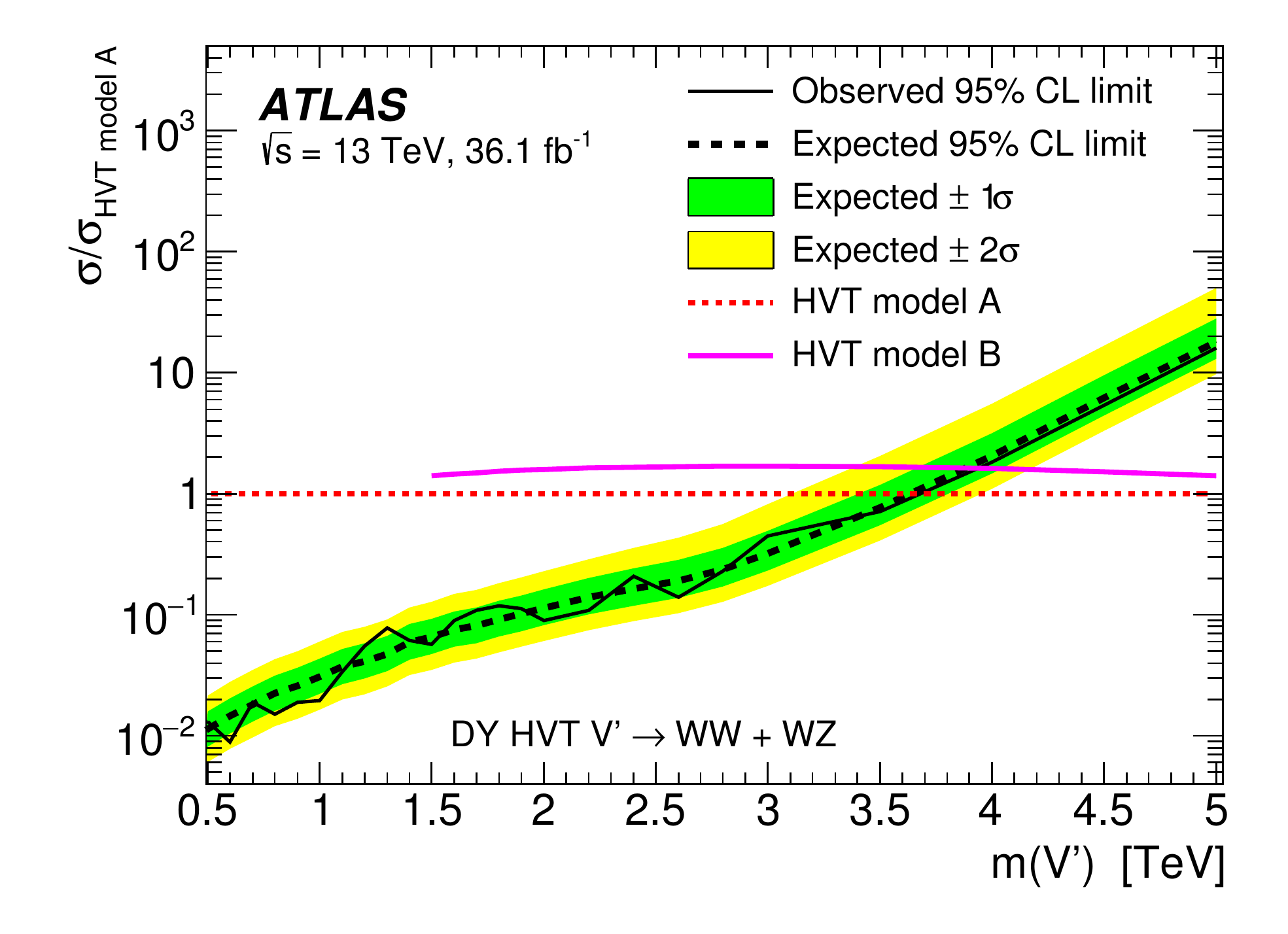}
    }
    \subfloat[]{
      \includegraphics[width=0.485\textwidth]{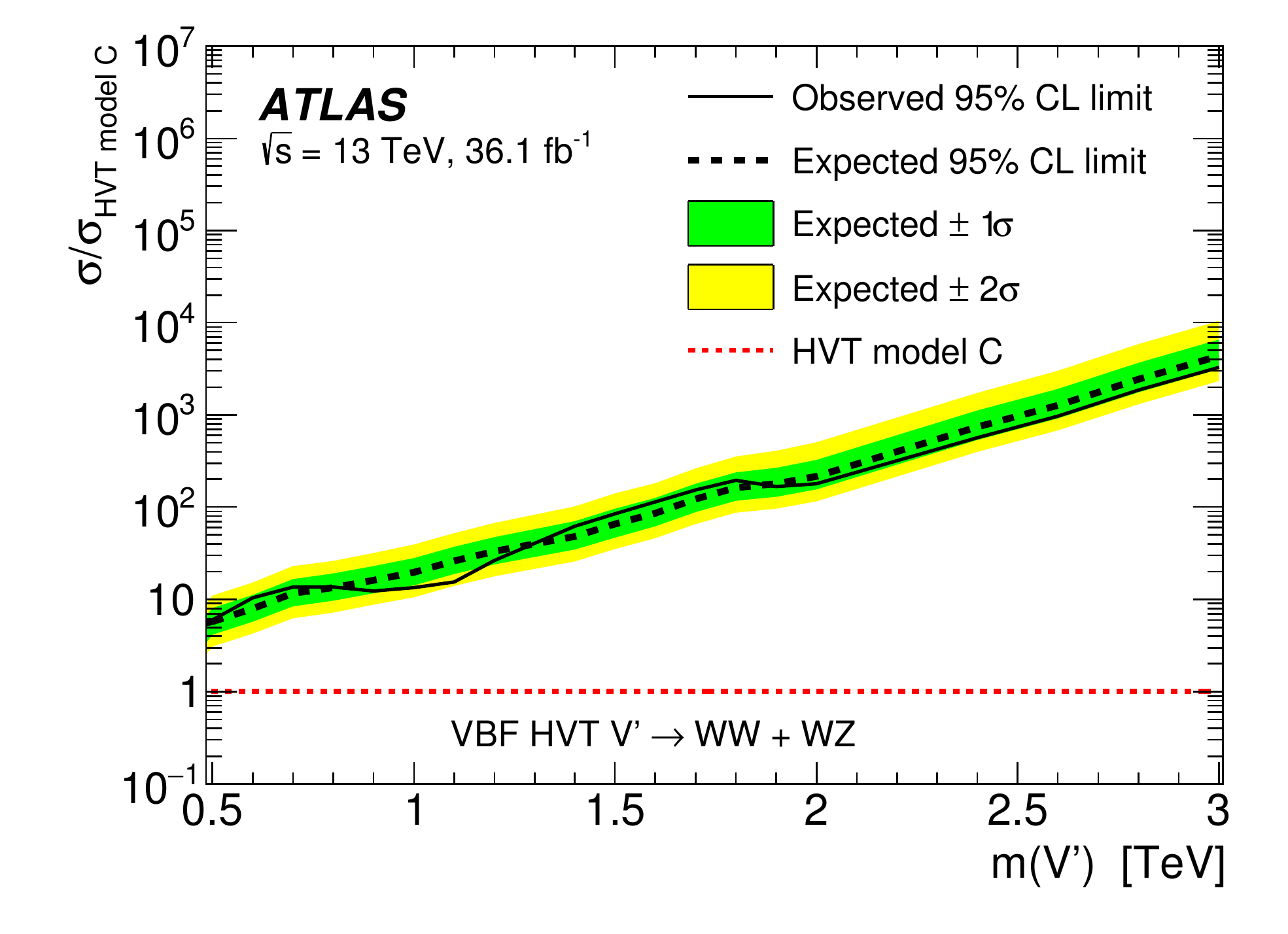}
    }
  \end{center}
  \vspace{-20 pt}
  \caption{Observed and expected 95\% CL upper limits on the $V^\prime$ cross section times branching fraction to $WW$ or $WZ$ for the HVT benchmark model, relative to the cross section times branching fraction for HVT model A or C, as applicable. Results are shown for (a) DY and (b) VBF production mechanisms. The model predictions are also shown.}
  \label{fig:hvtLims_WWorWZ}
\end{figure}

\begin{figure}[tb]
  \begin{center}
    \includegraphics[width=0.485\textwidth]{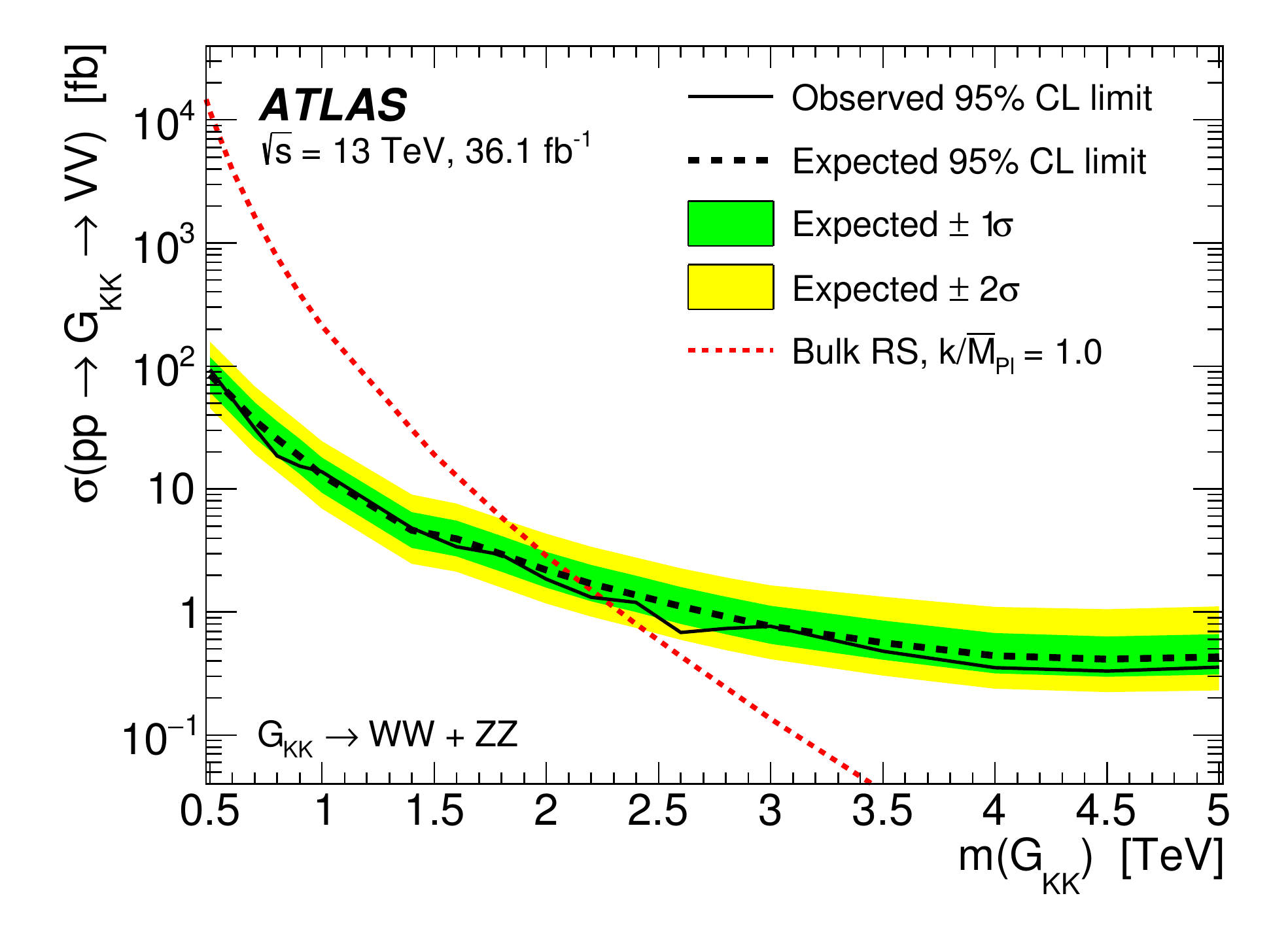}
  \end{center}
  \vspace{-20 pt}
  \caption{Observed and expected 95\% CL upper limits on the $\GKK$ cross section times branching fraction to $WW$ or $ZZ$ for the bulk RS model with $\kMPl = 1$. The model prediction is also shown.}
  \label{fig:rsgLims_WWorZZ}
\end{figure}

\begin{figure}[tb]
  \begin{center}
    \subfloat[]{
      \includegraphics[width=0.485\textwidth]{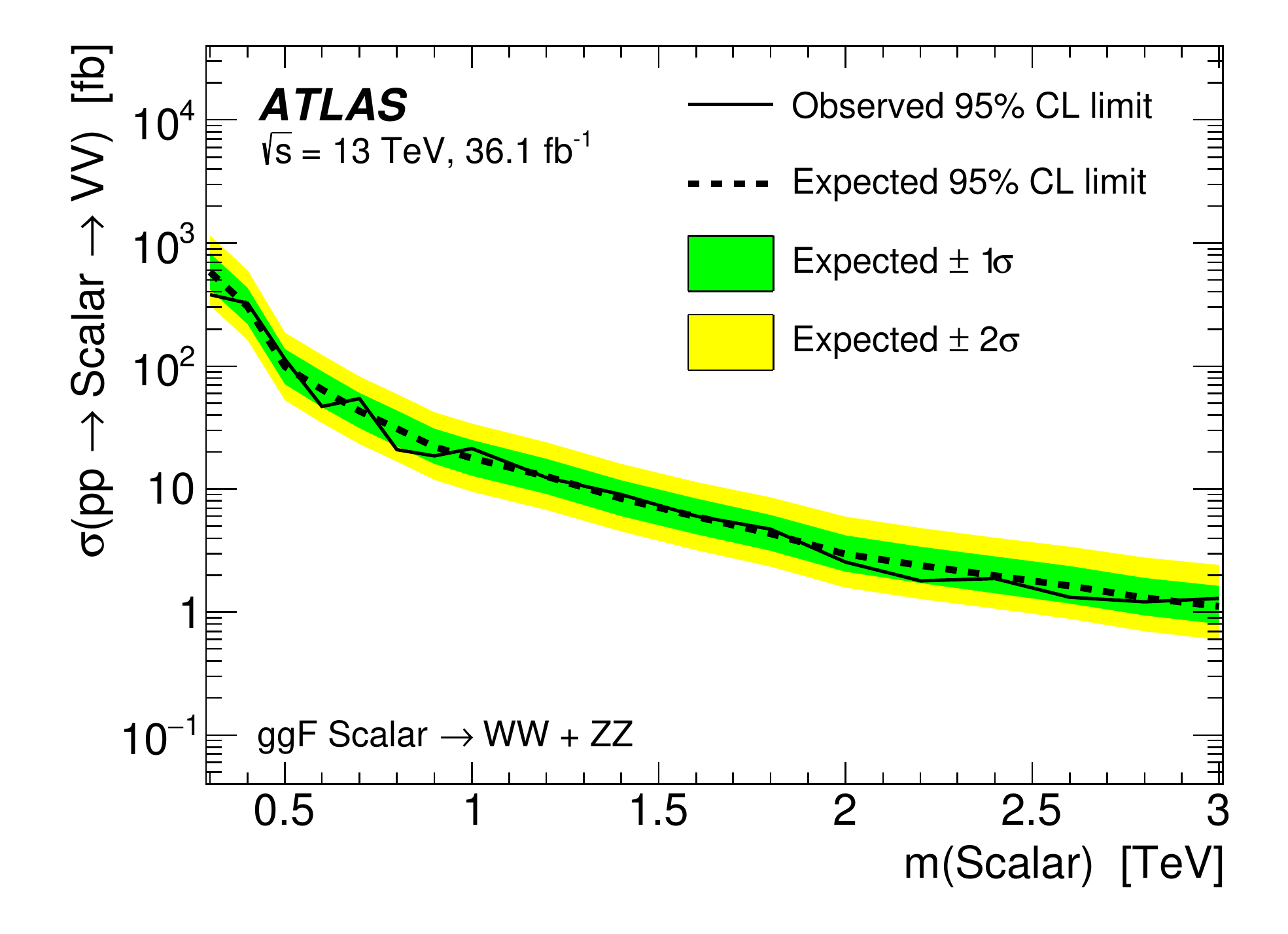}
    }
    \subfloat[]{
      \includegraphics[width=0.485\textwidth]{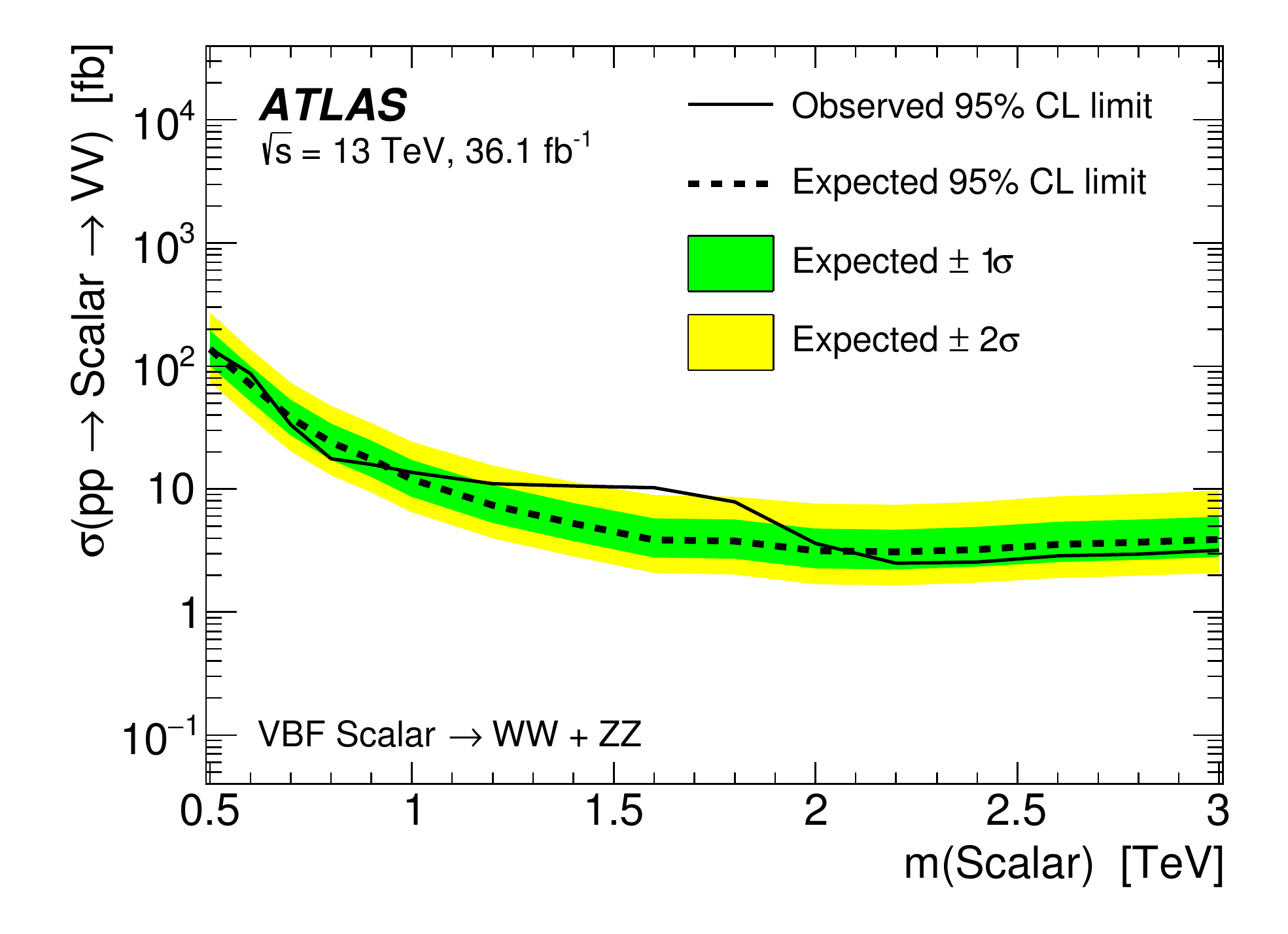}
    }
  \end{center}
  \vspace{-20 pt}
  \caption{Observed and expected 95\% CL upper limits on the cross section times branching fraction to $WW$ or $ZZ$ for a scalar particle. Results are shown for (a) DY and (b) VBF production mechanisms.}
  \label{fig:scalarLims_WWorZZ}
\end{figure}

For the combination of $VH$ search channels, only the HVT benchmark models are considered. Figure~\ref{fig:Lims_VH} shows the \sigbr\ limits relative to the HVT model A cross section, for decays into $WH$ and $ZH$ combined. 

\begin{figure}[tb]
  \begin{center}
    \includegraphics[width=0.485\textwidth]{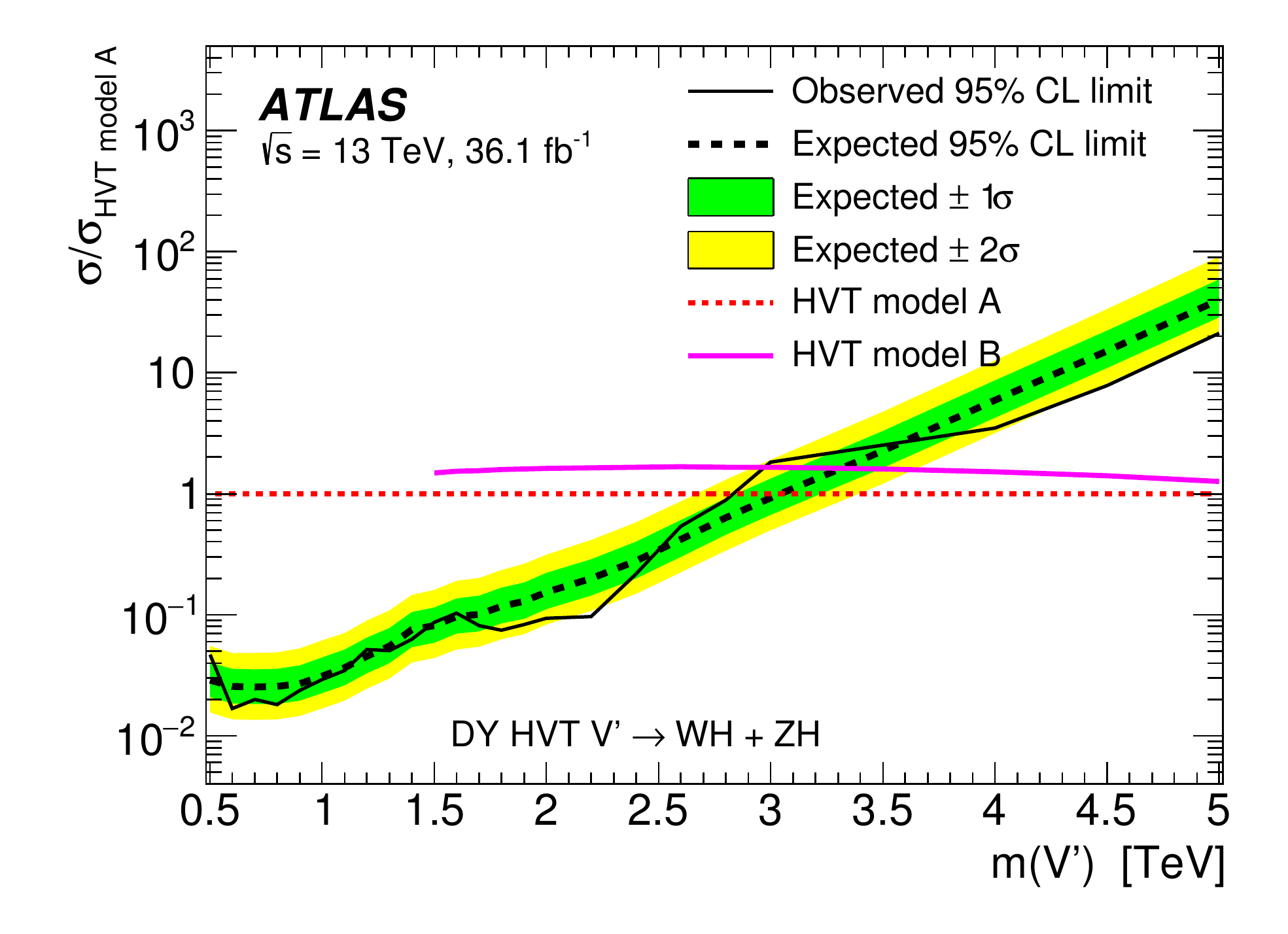}
  \end{center}
  \vspace{-20 pt}
  \caption{Observed and expected 95\% CL upper limits on the $V^\prime$ cross section times branching fraction to $WH$ or $ZH$ for the HVT benchmark model, relative to the cross section times branching fraction for HVT model A. The model predictions are also shown.}
  \label{fig:Lims_VH}
\end{figure}

The $VV$ and $VH$ channels are then combined, setting limits on \sigbr\ relative to the HVT model A prediction, as shown in Figure~\ref{fig:Lims_VVVH_lllv}(a). For the leptonic channels ($W' \rightarrow \ell\nu$ and $Z' \rightarrow \ell\ell$) only HVT model A is considered as shown in Figure~\ref{fig:Lims_VVVH_lllv}(b).

\begin{figure}[tb]
  \begin{center}
    \subfloat[]{
      \includegraphics[width=0.485\textwidth]{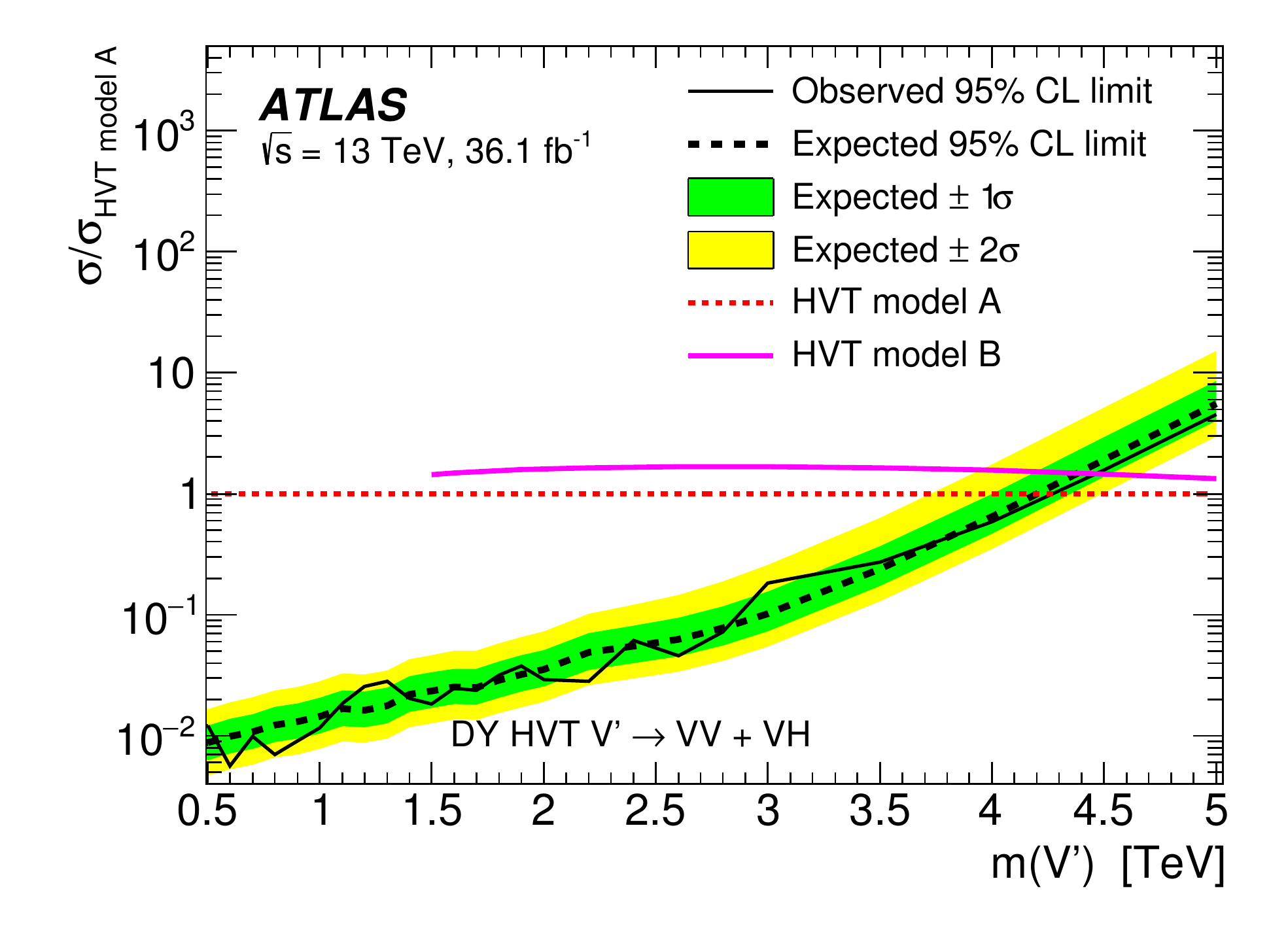}
    }
    \subfloat[]{
    \includegraphics[width=0.485\textwidth]{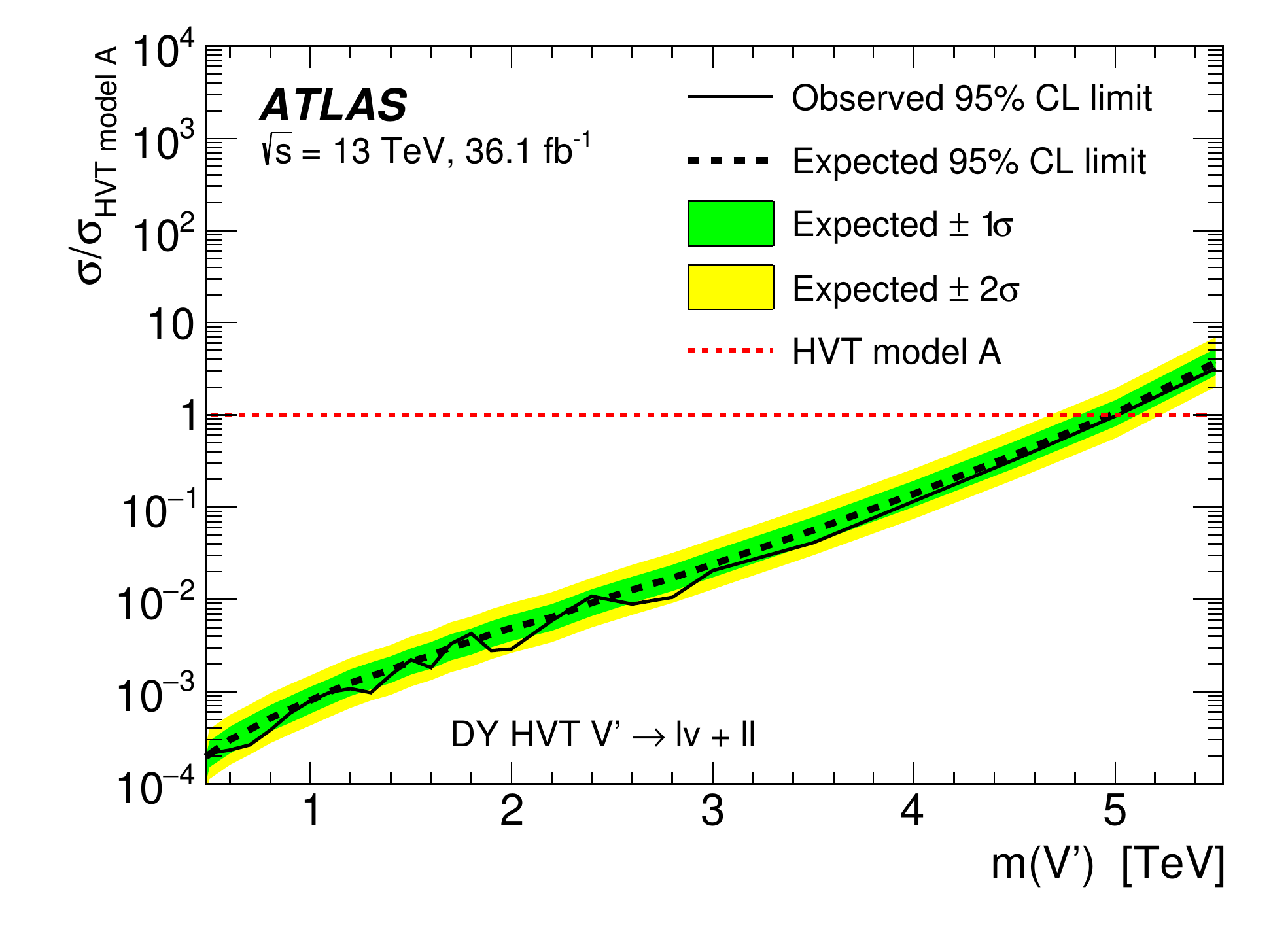}
    }
  \caption{Observed and expected 95\% CL upper limits on the $V^\prime$ cross section times branching fraction to (a) $VV/VH$ and (b) $\ell\nu$/$\ell\ell$ for the HVT benchmark model, relative to the cross section for HVT model A. The model predictions are also shown.}
  \label{fig:Lims_VVVH_lllv}
   \end{center}
\end{figure}

The channels are then further combined to set limits on HVT model A using not only $VV$ and $VH$ decay modes, but also $\ell\nu/\ell\ell$ decay modes. Figure~\ref{fig:Lims_All} presents the resulting limits on \sigbr\ relative to the HVT model A prediction for \wp, \zp, and $V^{\prime}$ production. Separate $VV/VH$ and $\ell\nu/\ell\ell$ expected limits are shown in Figure~\ref{fig:Lims_All}(d). As the $VV$ and $VH$ analyses only usually consider $V^{\prime}$ masses up to 5~\TeV, the acceptance is extrapolated to 5.5~\TeV\ for the full combination.

\begin{figure}[tb]
  \centering
  \subfloat[]{\includegraphics[width=0.485\textwidth]{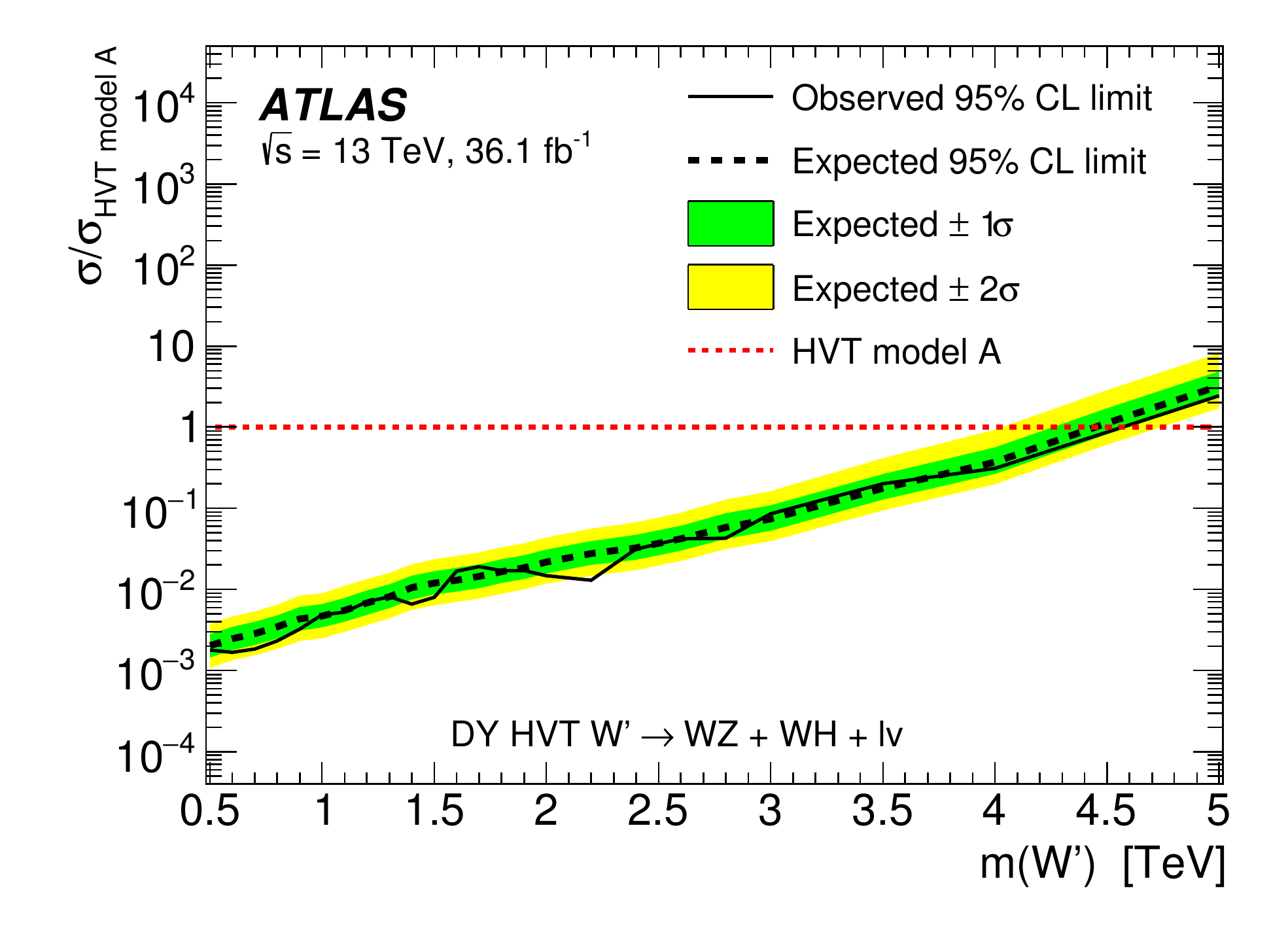}}
  \subfloat[]{\includegraphics[width=0.485\textwidth]{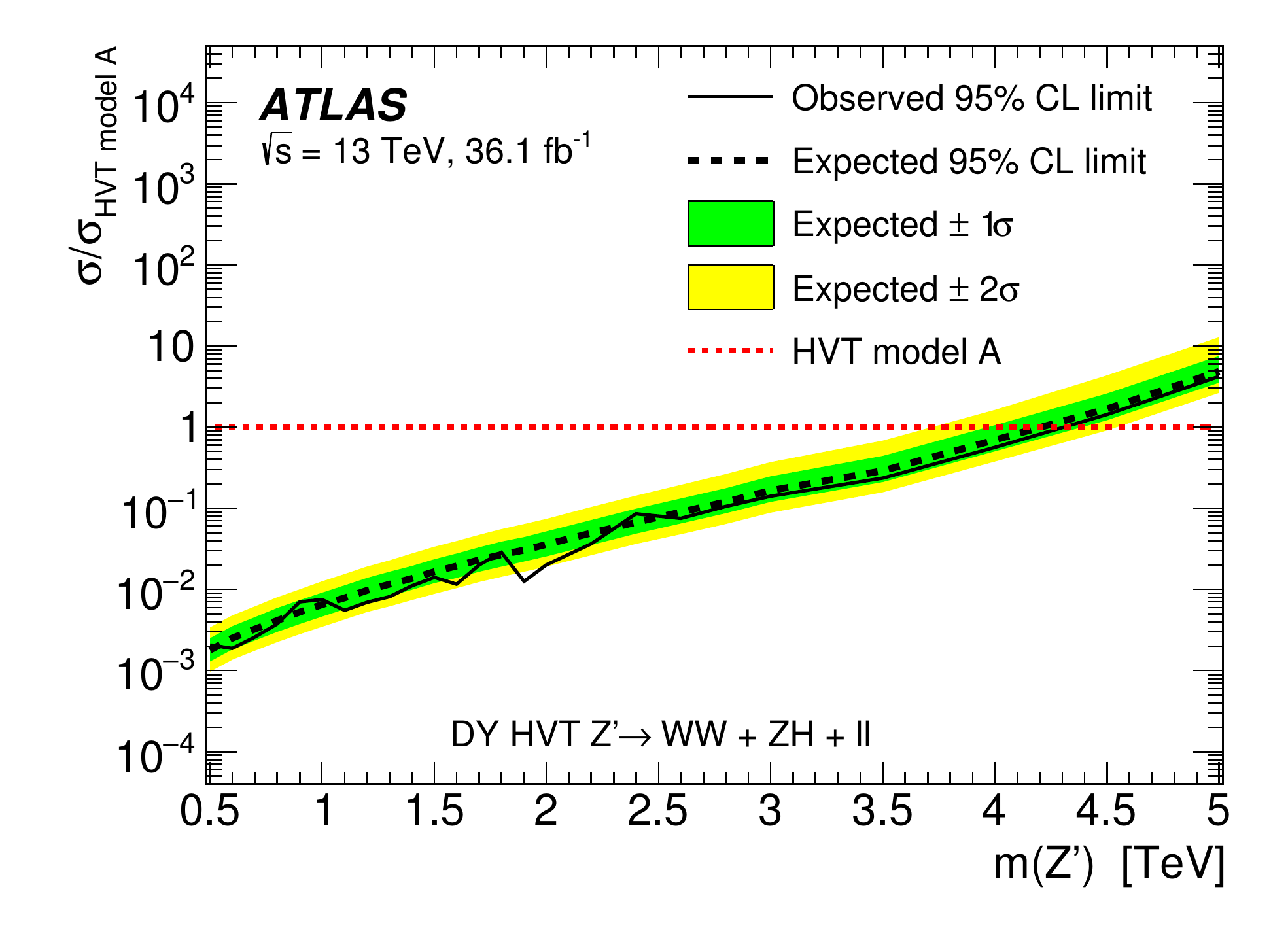}}\\
  \subfloat[]{\includegraphics[width=0.485\textwidth]{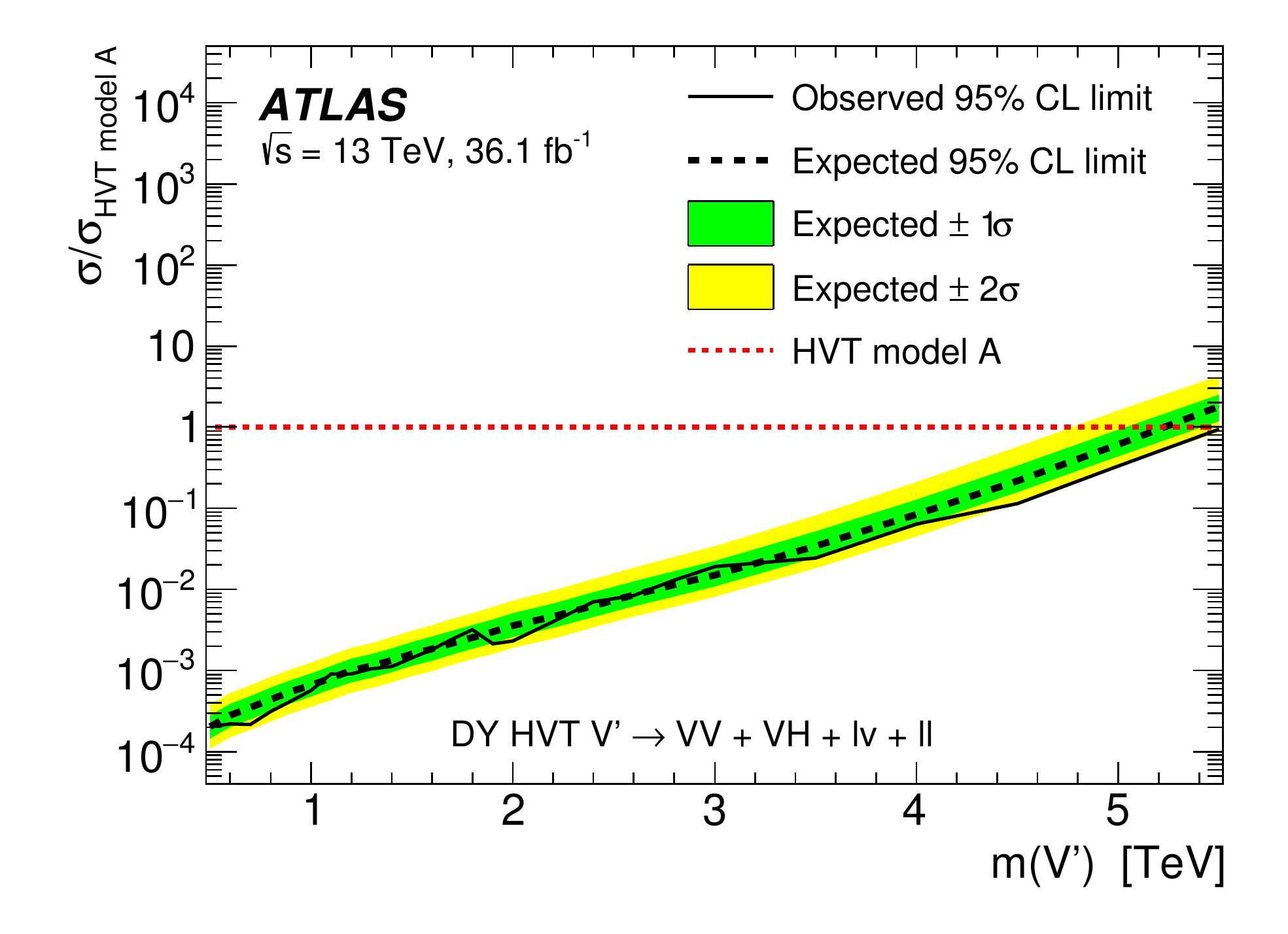}}
  \subfloat[]{\includegraphics[width=0.485\textwidth]{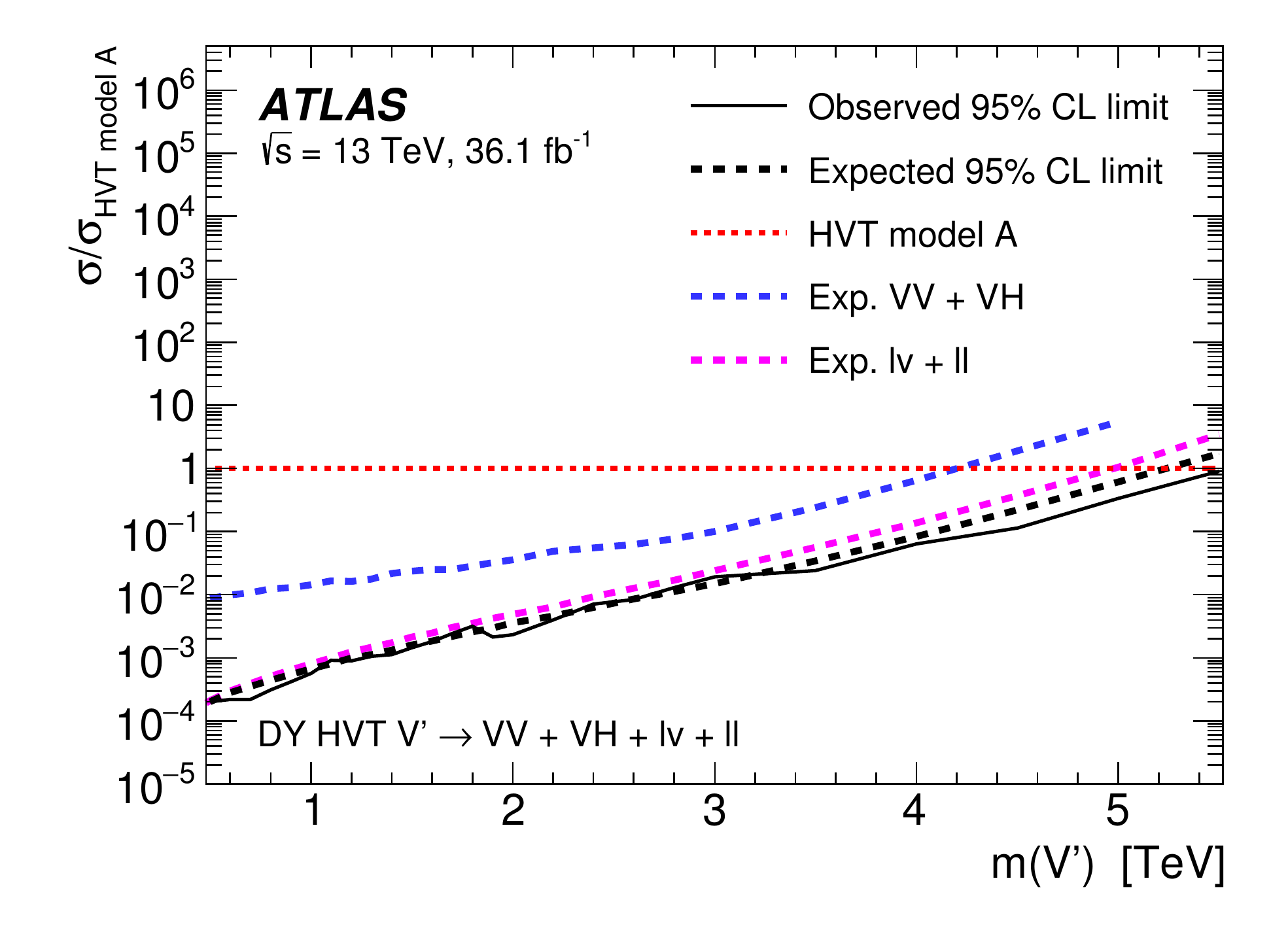}}
  \caption{Observed and expected 95\% CL upper limits on the $V^\prime$ cross section times branching fraction to $VV$, $VH$, and lepton--antilepton, relative to the prediction for HVT model A. Results are shown for (a) \wp, (b) \zp, and (c) $V^{\prime}$ production; (d) shows expected limits for bosonic and leptonic decay modes. The model predictions are also shown.}
\label{fig:Lims_All}
\end{figure}

Each of the channels presented here contributes uniquely to the search for heavy resonances, and the results obtained by their combination extend the reach beyond that of the individual searches. By using the HVT, bulk RS, and scalar benchmark models for comparison, the relative exclusion power of each search and their combinations can be compared. The intersection of the benchmark model predictions and the \sigbr\ upper limits yields lower limits on the resonance mass in each case.  The observed and expected lower limits on the resonance mass are summarized in Table~\ref{tab:MassLimits}.

\begin{table}[ht]
\caption{Observed and expected 95\% CL lower limits on resonance mass for benchmark models in each of the combined searches.}
\begin{center}
\begin{tabular}{l|cc|cc|cc}
\hline\hline
\multirow{3}{*}{Channel} & \multicolumn{6}{c}{Lower limits on resonance mass [TeV]} \\
\cline{2-7}
& \multicolumn{2}{|c}{HVT model A} & \multicolumn{2}{|c}{HVT model B} & \multicolumn{2}{|c}{Bulk RS} \\
\cline{2-7}
& Obs & Exp & Obs & Exp & Obs & Exp \\
\hline
$WW$ & 2.9 & 3.1 & 3.6 & 3.5 & 1.7 & 1.9 \\
$WZ$ & 3.6 & 3.6 & 3.9 & 3.9 & - & - \\
$ZZ$ & - & - & - & - & 1.5 & 1.7 \\
$VV$ & 3.7 & 3.7 & 4.0 & 3.9 & 2.3 & 2.2 \\
\hline
$WH$ & 2.6 & 2.8 & 2.8 & 3.1 & - & - \\
$ZH$ & 2.7 & 2.5 & 2.8 & 2.8 & - & - \\
$VH$ & 2.8 & 3.1 & 3.0 & 3.4 & - & - \\
\hline
$\ell\nu$ & 4.6 & 4.6 & - & - & - & - \\
$\ell\ell$ & 4.5 & 4.4 & - & - & - & - \\
$\ell\nu/\ell\ell$ & 5.0 & 5.0 & - & - & - & - \\
\hline
$VV/VH$ & 4.3 & 4.3 & 4.5 & 4.4 & - & - \\
$VV/VH/\ell\nu/\ell\ell$ & 5.5 & 5.3 & - & - & - & - \\
\hline\hline
\end{tabular}
\label{tab:MassLimits}
\end{center}
\end{table}

\clearpage

The search channels included here provide access to several coupling strengths of heavy resonances to SM particles as described by Eq.~(\ref{eq:lagrangian}) in the context of the HVT model. Specifically, the data constrain the coupling strength to both the quarks and bosons in the $VV$ and $VH$ channels, whereas constraints are placed on both the quark and lepton couplings in the leptonic channels. These constraints are shown in Figures~\ref{fig:HVTcouplingsI},~\ref{fig:HVTcouplingsII_inclEW},~and~\ref{fig:HVTcouplingsIII_inclEW}, where the first and second include a shaded area denoting a region where the limits are not valid because resonances would have a width greater than 5\% of their mass. This is a region where the resonance width would exceed the discriminating variable's resolution in the search, and the assumed narrow-width approximation breaks down. Figures~\ref{fig:HVTcouplingsII_inclEW}~and~\ref{fig:HVTcouplingsIII_inclEW} include constraints on heavy resonances with masses of 3, 4, and 5~\TeV\ from precision electroweak (EW) measurements~\cite{delAguila:2010mx}, which already exclude this aforementioned region for the relevant contours shown. The EW constraints are only overlaid on the final plots for each part of the combination.

The constraints from the $VV$, $VH$, and combined $VV$ and $VH$ channels are presented in Figure~\ref{fig:HVTcouplingsI}, showing the $\{g_H, g_f\}$ plane for each as well as the $\{g_q, g_\ell\}$ plane for $VV$/$VH$. These constraints are strongest at large couplings for both $g_f$ and $g_H$ but become weak as these couplings approach zero. This is because the resonance couplings to $VV$ and $VH$ tend to zero as the $g_H$ coupling approaches zero, and production of the resonance also tends to zero as the $g_f$ coupling approaches zero. The constraints in the $\{g_q, g_\ell\}$ plane shown in Figure~\ref{fig:HVTcouplingsI}(d) weaken at larger $|g_\ell|$ values due to an increase in the leptonic branching fraction and a corresponding decrease in the bosonic branching fraction.

Figure~\ref{fig:HVTcouplingsII_inclEW} presents the constraints from the $\ell\nu/\ell\ell$ channels in the $\{g_H, g_f\}$ plane, $\{g_q, g_\ell\}$ plane for $g_H$ set to the value from HVT model A, and $\{g_q, g_\ell\}$ plane for $g_H$ set to 0. In this last case the bosonic channels do not contribute because $g_H$ = 0, meaning only the leptonic channels contribute. As the leptonic channels involve direct production of a $V^{\prime}$ resonance and subsequent decay, without intermediate bosons, the constraints remain strong even as $g_H$ tends to zero, and in fact are strongest there due to the restriction of alternative decay modes. The constraints from these channels still weaken as the $g_f$ coupling tends to zero though, as it does when $g_\ell$ and/or $g_q$ tends to zero. These features demonstrate the complementarity between the $VV/VH$ and $\ell\nu/\ell\ell$ decay modes. 

The complementarity is further evidenced by the full $VV$/$VH$/$\ell\nu$/$\ell\ell$ combination in both the $\{g_H, g_f\}$ plane, as shown in Figure~\ref{fig:HVTcouplingsIII_inclEW}(a), and the $\{g_q, g_\ell\}$ plane, as shown in Figure~\ref{fig:HVTcouplingsIII_inclEW}(b). The resulting constraints are very stringent, improving on the limits from current precision EW measurements in almost all areas of the respective planes, except at low $|g_q|$ values when considering non-universal quark and lepton couplings in the $\{g_q, g_\ell\}$ plane. This is due to the asymmetry of the precision EW measurement limits, which is related to interference effects. The constraints for HVT model A are generally stronger than for model B, due to the small fermion couplings in the latter scenario.

\begin{figure}[tb]
	\centering
	\subfloat[]{
      		\includegraphics[width=0.485\textwidth]{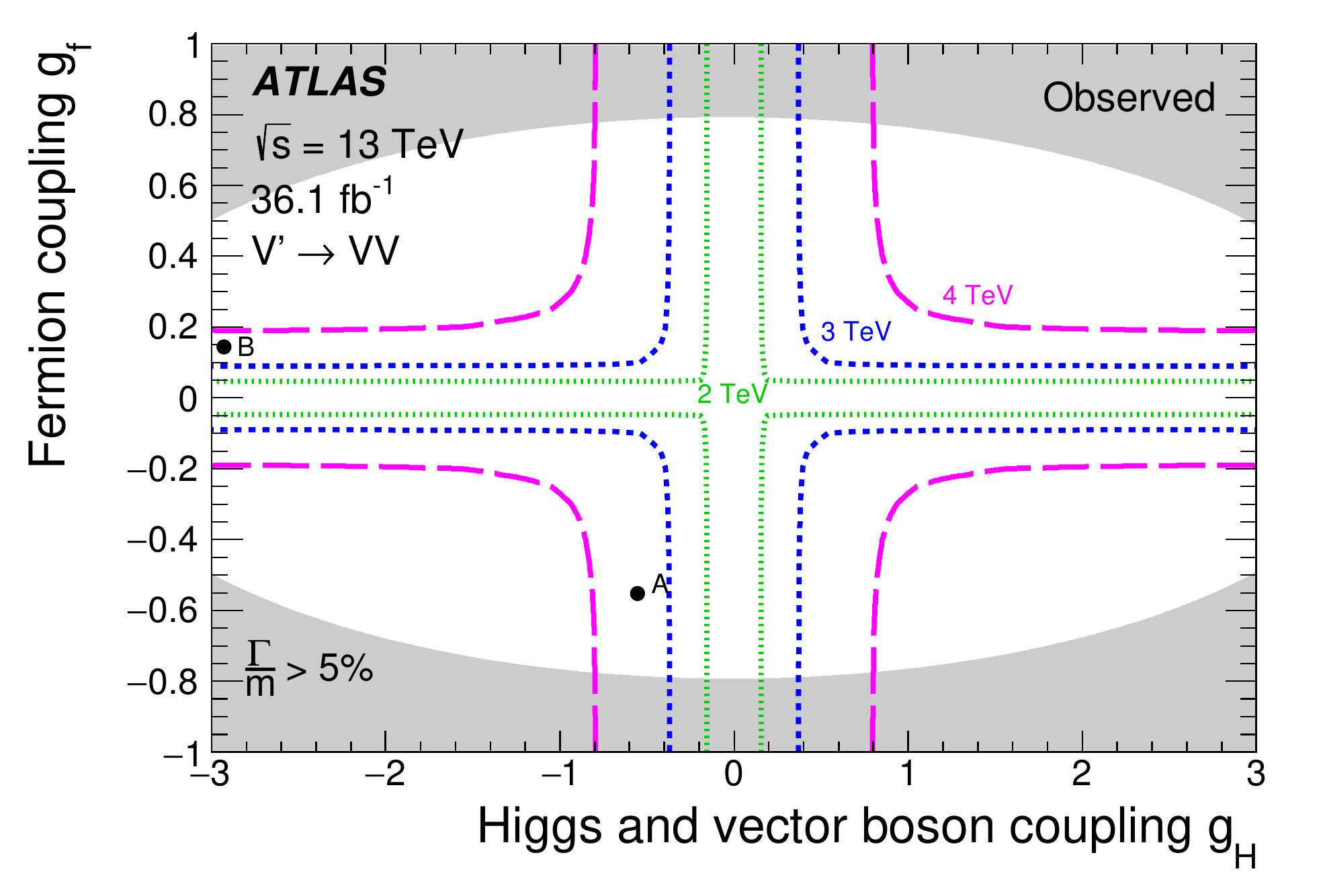}
      	}
        \subfloat[]{
      		\includegraphics[width=0.485\textwidth]{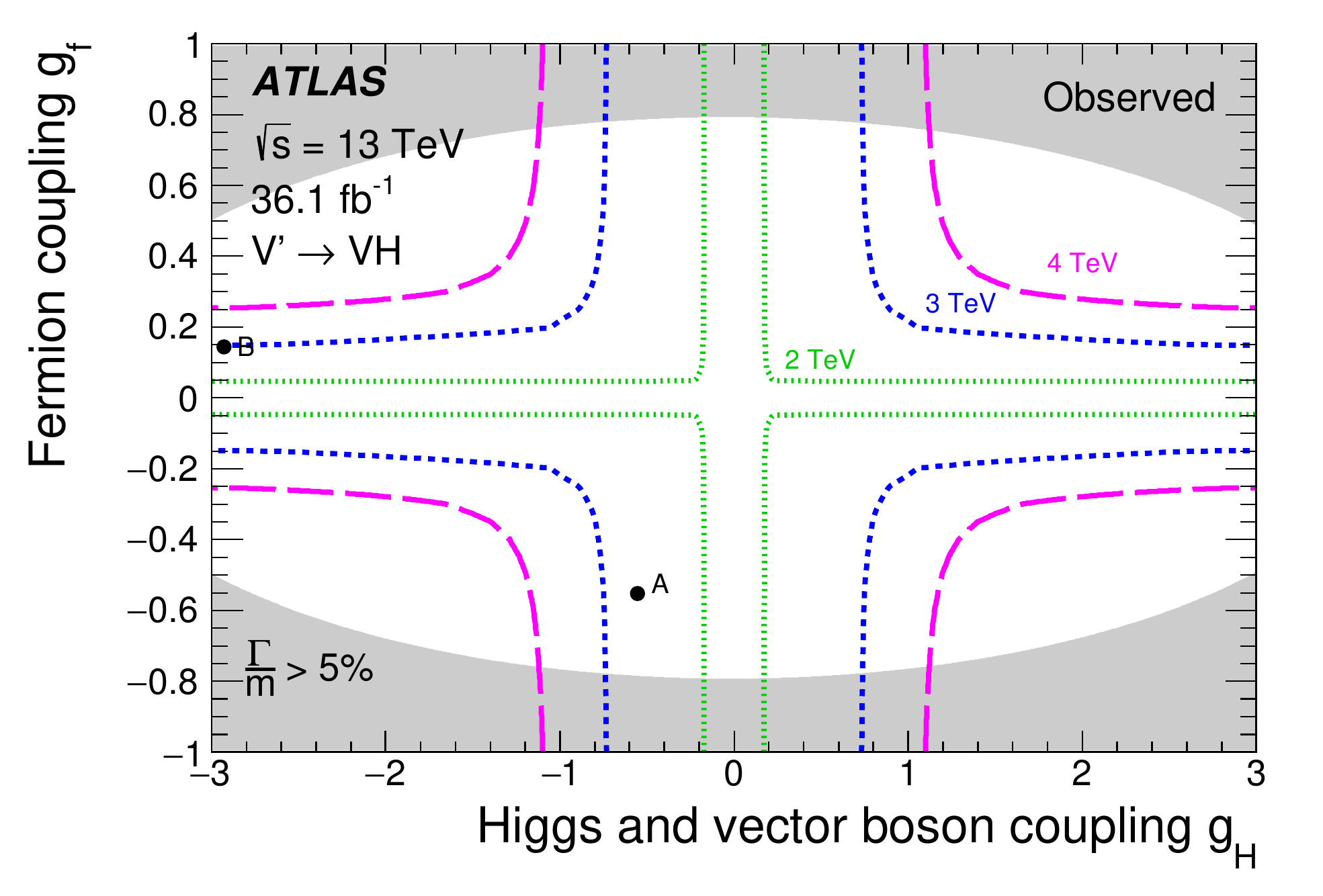}
      	}\\
	\subfloat[]{
      		\includegraphics[width=0.485\textwidth]{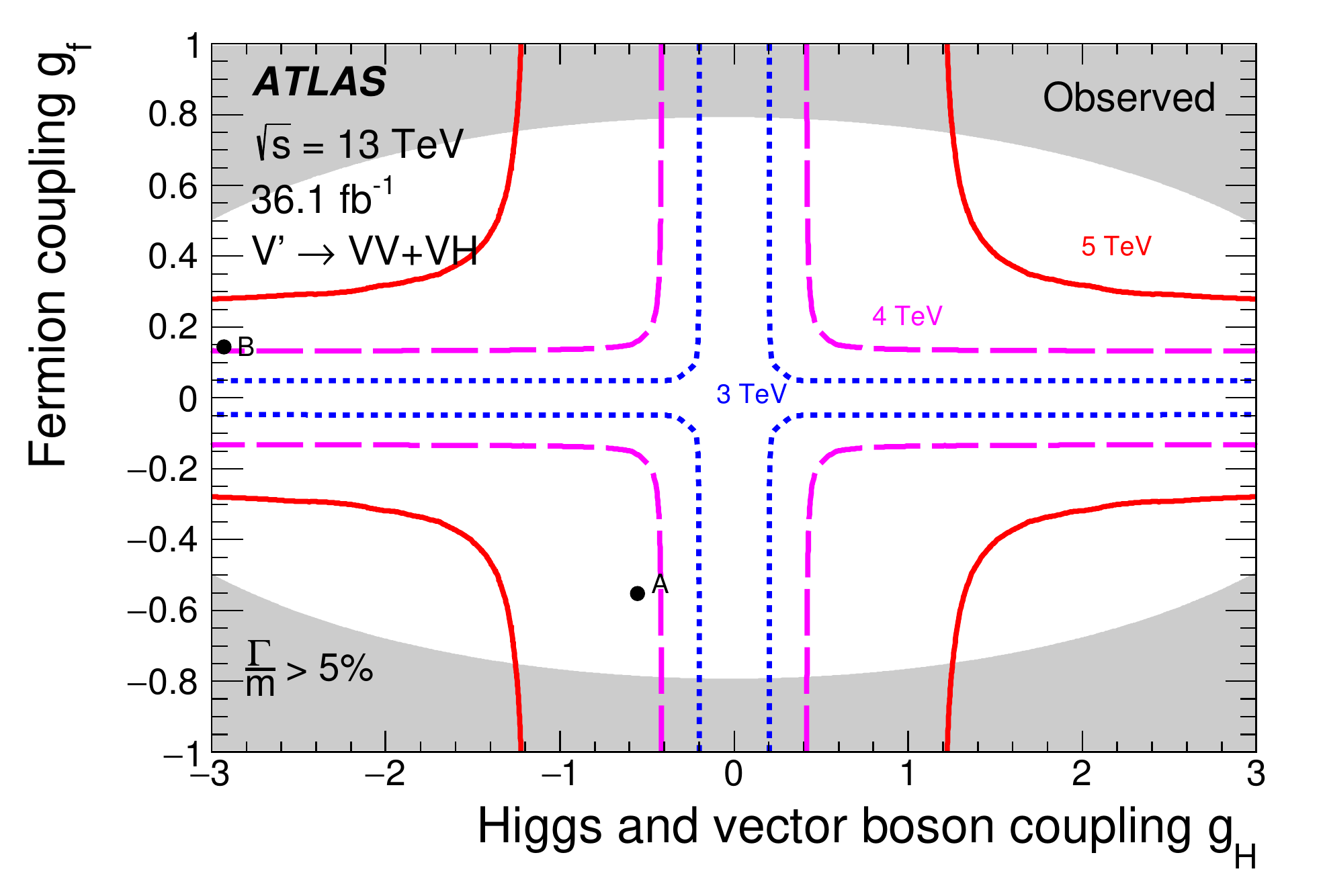}
      	}
        \subfloat[]{
      		\includegraphics[width=0.485\textwidth]{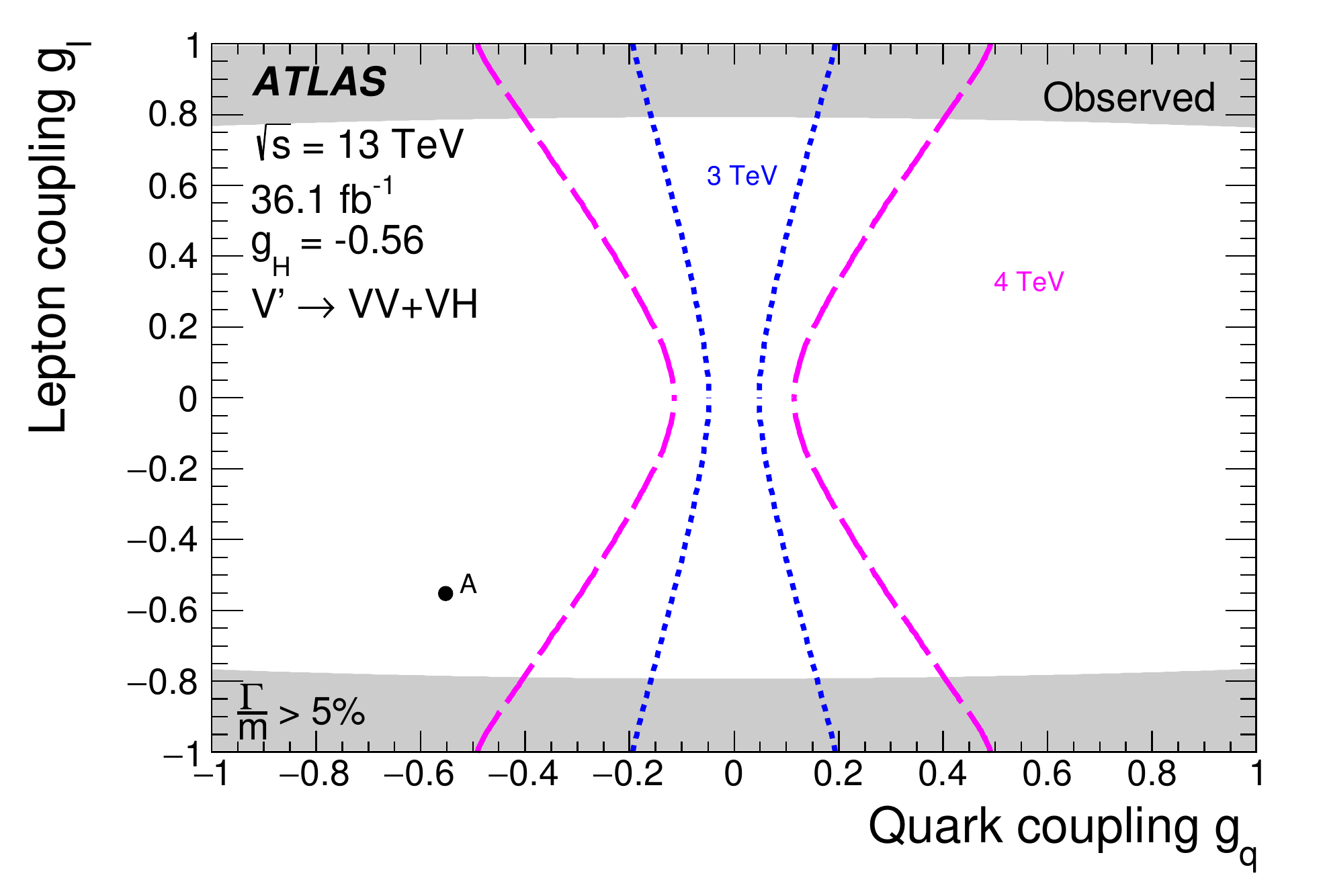}
      	}
  \caption{Observed 95\% CL exclusion contours in the $\{g_H, g_f\}$ HVT parameter space for resonances
   of mass 2, 3, 4, or 5~\TeV\ for (a) the $VV$ channels, (b) the $VH$ channels, and
   (c) the combined $VV$ and $VH$ channels; (d) shows the combined $VV$ and $VH$ channels in the $\{g_q, g_\ell\}$ plane with $g_H$ set to the value from HVT model A. For (d) the 5~\TeV\ limit contour is outside of the plane. The areas outside the curves are excluded. The gray area indicates parameter regions for which $\Gamma/m > 5\%$. Also shown are the parameters for models A and B, where applicable.}
\label{fig:HVTcouplingsI}
\end{figure}

\begin{figure}[tb]
	\centering
	\subfloat[]{
      		\includegraphics[width=0.485\textwidth]{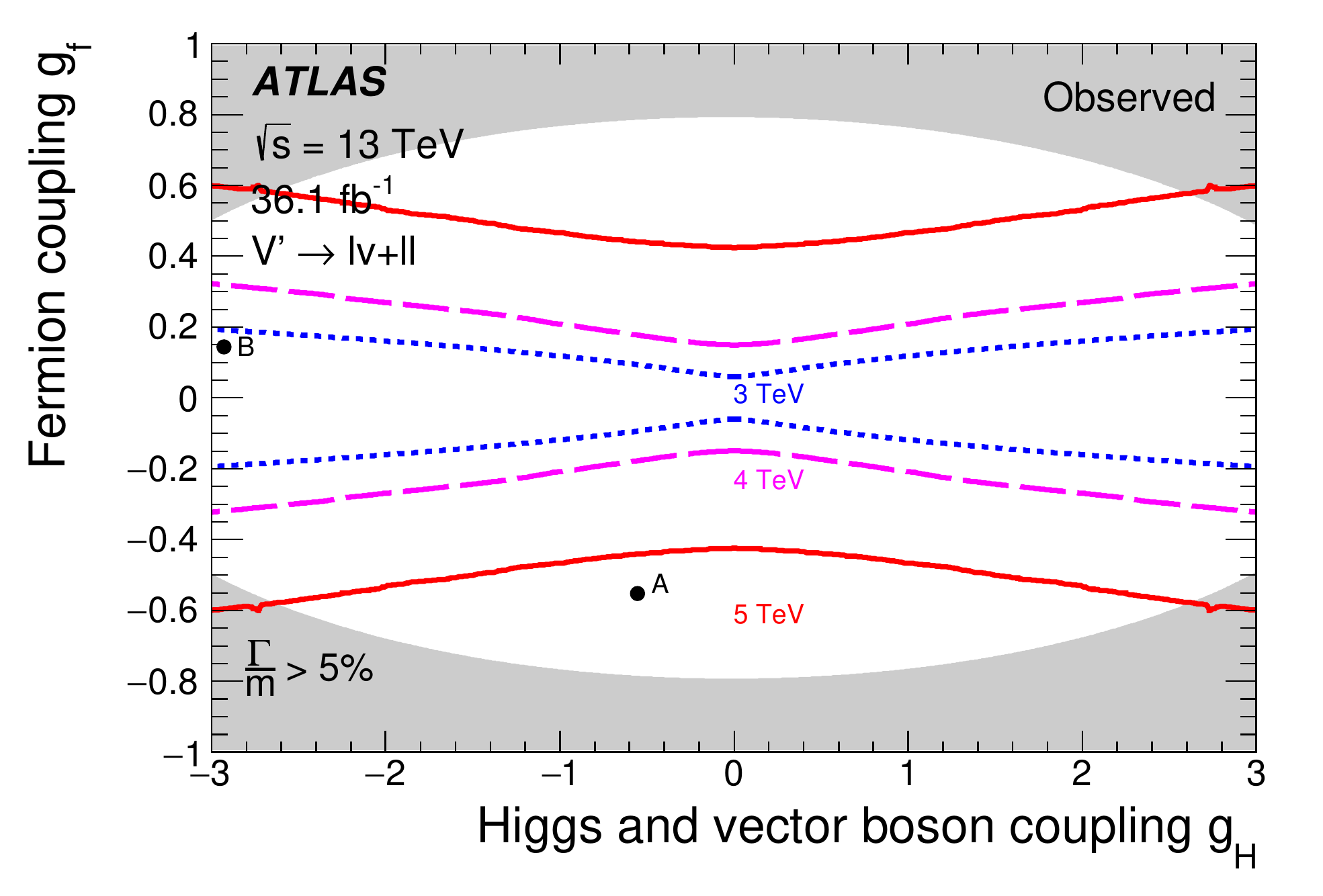}
      	}\\
        \subfloat[]{
      		\includegraphics[width=0.485\textwidth]{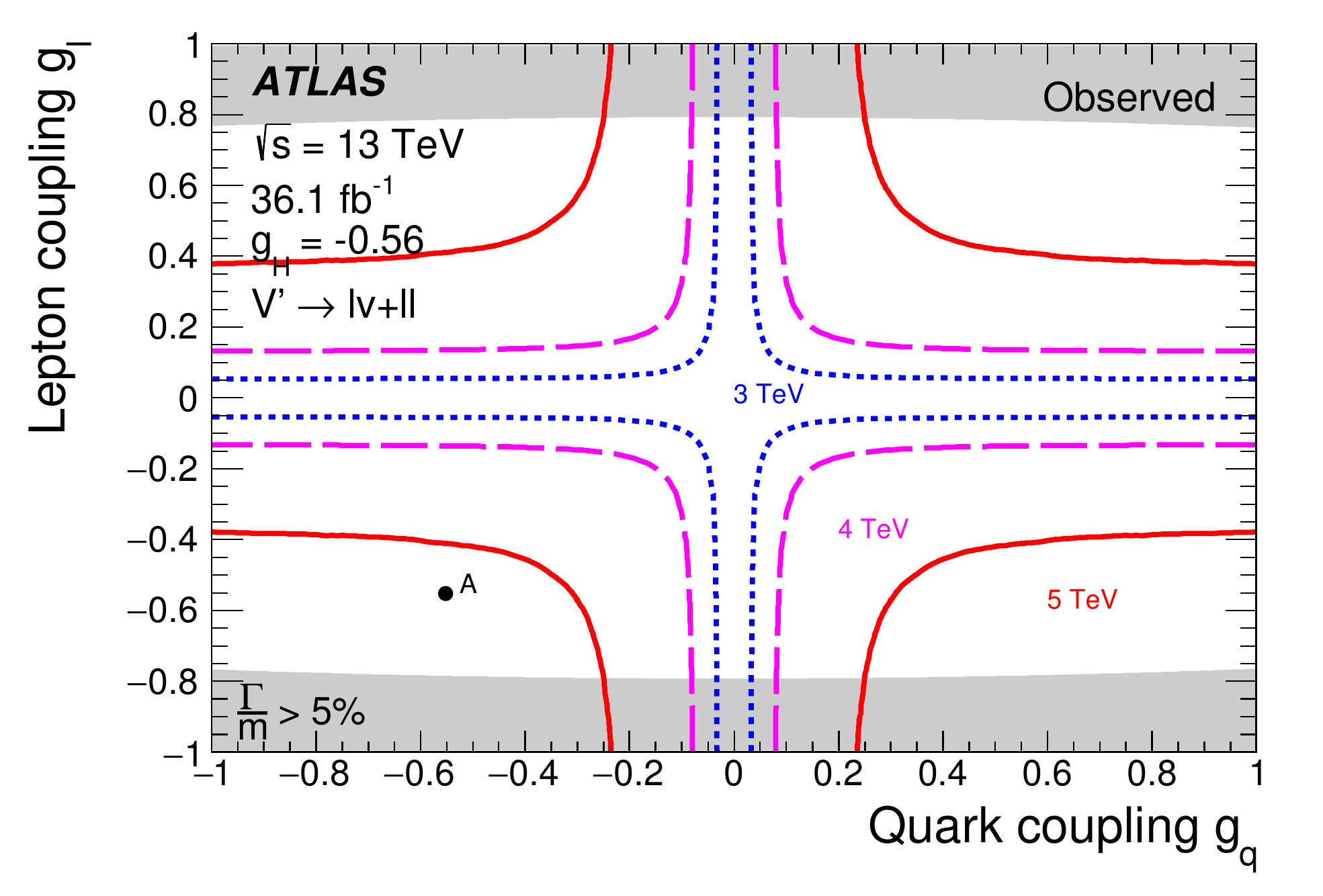}
      	}
        \subfloat[]{
      		\includegraphics[width=0.485\textwidth]{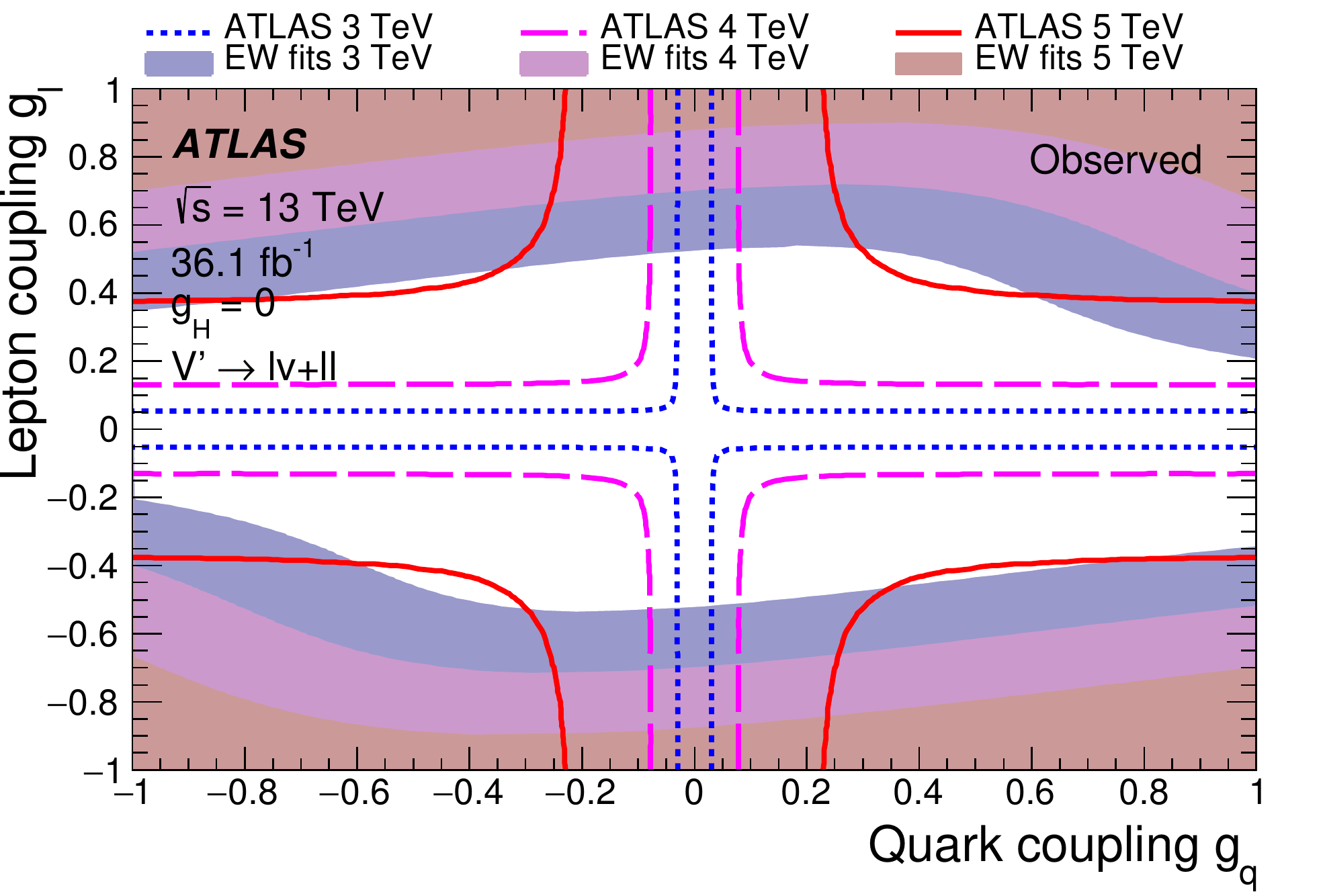}
      	}
  \caption{Observed 95\% CL exclusion contours in the HVT parameter space (a) $\{g_H, g_f\}$,
   (b) $\{g_q, g_\ell\}$ with $g_H$ set to the value from HVT model A, and (c) $\{g_q, g_\ell\}$ with $g_H$ set to 0, for resonances of mass 3, 4, and 5~\TeV\ for the $\ell\nu/\ell\ell$
   channels. The areas outside the curves are excluded, as are the filled regions which show the constraints from precision EW measurements.
   The gray area indicates parameter regions for which $\Gamma/m > 5\%$. 
   Also shown are the parameters for models A and B, where applicable.}
\label{fig:HVTcouplingsII_inclEW}
\end{figure}

\begin{figure}[tb]
	\centering
	\subfloat[]{
      		\includegraphics[width=0.485\textwidth]{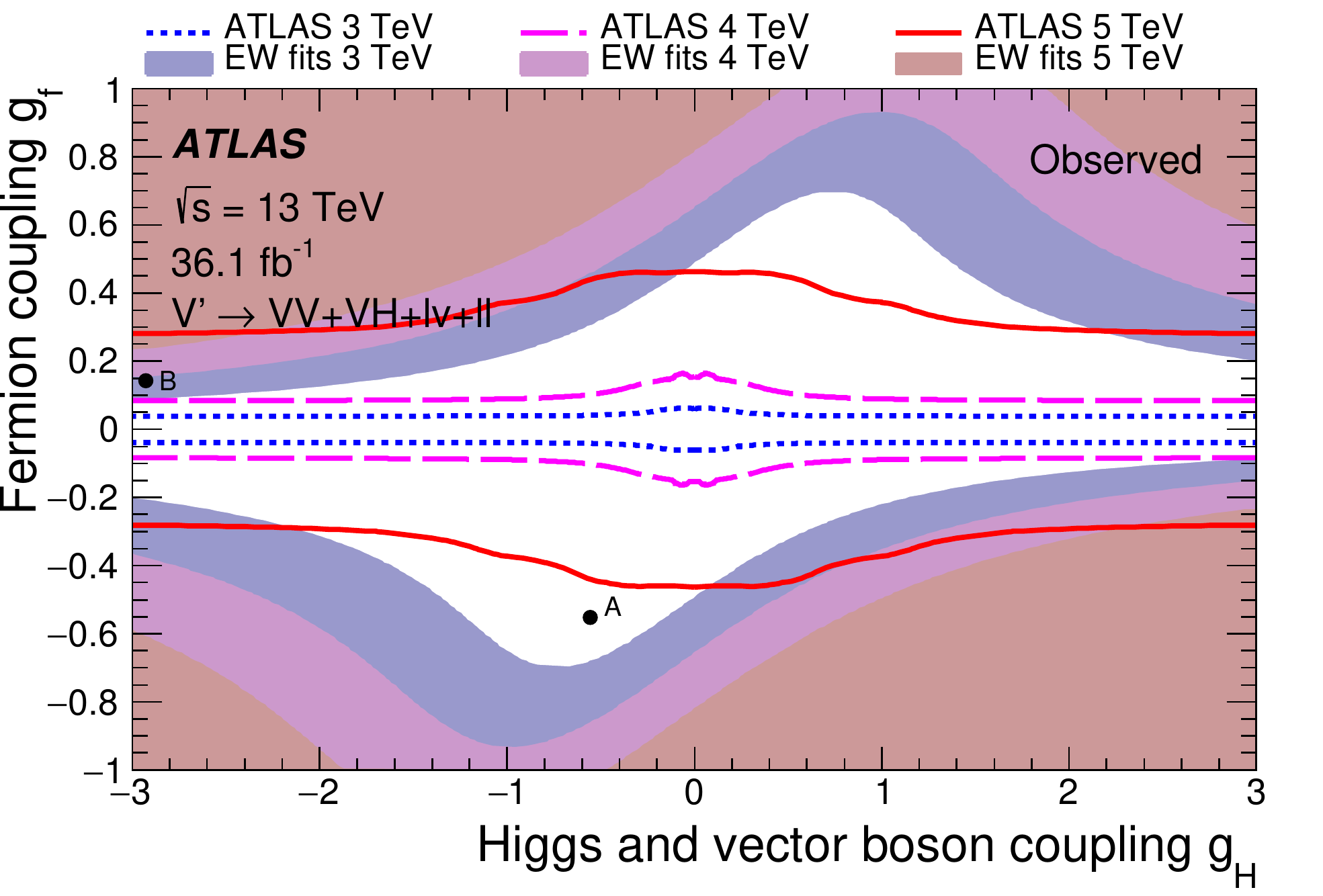}
      	}
      \subfloat[]{
      		\includegraphics[width=0.485\textwidth]{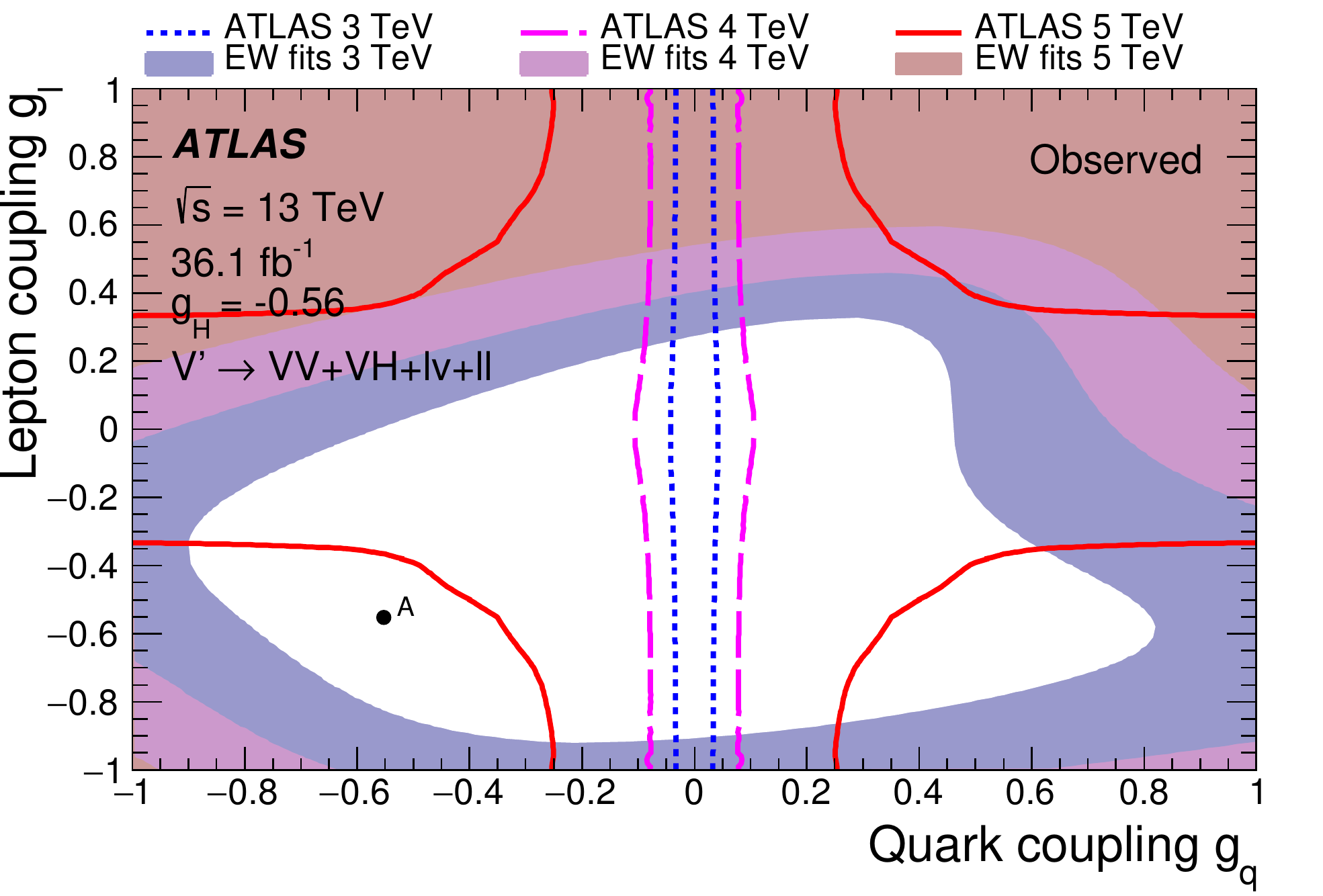}
      	}
  \caption{Observed 95\% CL exclusion contours in the HVT parameter space (a) $\{g_H, g_f\}$ and
   (b) $\{g_q, g_\ell\}$ for resonances of mass 3, 4, and 5~\TeV\ for the combination
   of the $VV$, $VH$, and $\ell\nu/\ell\ell$ channels. The areas outside the curves are excluded, as are the filled regions which show the constraints from precision EW measurements.
   Also shown are the parameters for models A and B, where applicable.}
\label{fig:HVTcouplingsIII_inclEW}
\end{figure}


\FloatBarrier
\section{Conclusions}
\label{sec:conclusion}
A combination of results from searches for heavy resonance production in various bosonic 
and leptonic final states is presented. The data were collected with the ATLAS detector at the LHC
in $pp$ collisions at $\rts = 13$~\TeV{} and correspond to an integrated luminosity of 36.1~\ifb.
While previous combination efforts included only the decays of heavy resonances into $VV$ and $VH$,
the combination presented here also includes decays into lepton--antilepton final states.
Compared to the individual analyses, the combined results strengthen the constraints on physics beyond the Standard Model and allow the constraints to be expressed in terms of the couplings
to quarks, leptons, or bosons. The relative sensitivities of the different
approaches are compared, including bosonic and leptonic final states or different production
mechanisms such as quark--antiquark annihilation/gluon--gluon fusion vs. vector-boson fusion.

The combined results are interpreted in the context of models with a heavy vector-boson triplet,
a Kaluza--Klein excitation of the graviton or a heavy scalar singlet.
The 95\% CL lower limit on the mass of $V^\prime$ resonances in the weakly coupled HVT model A 
is 5.5~\TeV\ 
and the corresponding limit in the strongly coupled HVT model B is 4.5~\TeV.
Similarly, the lower limit on the \GKK\ mass in the bulk RS model with $\kMPl = 1$ is 2.3~\TeV.
Limits on the cross section times branching fraction for an empirical heavy scalar model
range between 380 and 1.3~fb for scalar mass values between 0.3 and 3.0~\TeV\ in the case
of production via gluon--gluon fusion. The corresponding values for scalar production via
vector-boson fusion range between 140 and 3.2~fb for scalar masses between 0.5 and 3.0~\TeV.
Finally, the combined results are used to place stringent constraints on couplings of heavy
vector bosons to quarks, leptons, and bosons. 
Except at low values of quark couplings where resonance production via quark--antiquark annihilation 
is suppressed at the LHC, these constraints are found to be more stringent
than those extracted from precision electroweak measurements.

\section*{Acknowledgments}

We thank CERN for the very successful operation of the LHC, as well as the
support staff from our institutions without whom ATLAS could not be
operated efficiently.

We acknowledge the support of ANPCyT, Argentina; YerPhI, Armenia; ARC, Australia; BMWFW and FWF, Austria; ANAS, Azerbaijan; SSTC, Belarus; CNPq and FAPESP, Brazil; NSERC, NRC and CFI, Canada; CERN; CONICYT, Chile; CAS, MOST and NSFC, China; COLCIENCIAS, Colombia; MSMT CR, MPO CR and VSC CR, Czech Republic; DNRF and DNSRC, Denmark; IN2P3-CNRS, CEA-DRF/IRFU, France; SRNSFG, Georgia; BMBF, HGF, and MPG, Germany; GSRT, Greece; RGC, Hong Kong SAR, China; ISF, I-CORE and Benoziyo Center, Israel; INFN, Italy; MEXT and JSPS, Japan; CNRST, Morocco; NWO, Netherlands; RCN, Norway; MNiSW and NCN, Poland; FCT, Portugal; MNE/IFA, Romania; MES of Russia and NRC KI, Russian Federation; JINR; MESTD, Serbia; MSSR, Slovakia; ARRS and MIZ\v{S}, Slovenia; DST/NRF, South Africa; MINECO, Spain; SRC and Wallenberg Foundation, Sweden; SERI, SNSF and Cantons of Bern and Geneva, Switzerland; MOST, Taiwan; TAEK, Turkey; STFC, United Kingdom; DOE and NSF, United States of America. In addition, individual groups and members have received support from BCKDF, the Canada Council, CANARIE, CRC, Compute Canada, FQRNT, and the Ontario Innovation Trust, Canada; EPLANET, ERC, ERDF, FP7, Horizon 2020 and Marie Sk{\l}odowska-Curie Actions, European Union; Investissements d'Avenir Labex and Idex, ANR, R{\'e}gion Auvergne and Fondation Partager le Savoir, France; DFG and AvH Foundation, Germany; Herakleitos, Thales and Aristeia programmes co-financed by EU-ESF and the Greek NSRF; BSF, GIF and Minerva, Israel; BRF, Norway; CERCA Programme Generalitat de Catalunya, Generalitat Valenciana, Spain; the Royal Society and Leverhulme Trust, United Kingdom.

The crucial computing support from all WLCG partners is acknowledged gratefully, in particular from CERN, the ATLAS Tier-1 facilities at TRIUMF (Canada), NDGF (Denmark, Norway, Sweden), CC-IN2P3 (France), KIT/GridKA (Germany), INFN-CNAF (Italy), NL-T1 (Netherlands), PIC (Spain), ASGC (Taiwan), RAL (UK) and BNL (USA), the Tier-2 facilities worldwide and large non-WLCG resource providers. Major contributors of computing resources are listed in Ref.~\cite{ATL-GEN-PUB-2016-002}.



\printbibliography


\clearpage
\clearpage
 
\begin{flushleft}
{\Large The ATLAS Collaboration}

\bigskip

M.~Aaboud$^\textrm{\scriptsize 34d}$,    
G.~Aad$^\textrm{\scriptsize 99}$,    
B.~Abbott$^\textrm{\scriptsize 124}$,    
O.~Abdinov$^\textrm{\scriptsize 13,*}$,    
B.~Abeloos$^\textrm{\scriptsize 128}$,    
D.K.~Abhayasinghe$^\textrm{\scriptsize 91}$,    
S.H.~Abidi$^\textrm{\scriptsize 164}$,    
O.S.~AbouZeid$^\textrm{\scriptsize 39}$,    
N.L.~Abraham$^\textrm{\scriptsize 153}$,    
H.~Abramowicz$^\textrm{\scriptsize 158}$
% ...
\bigskip
\\

% ...
$^{13}$Institute of Physics, Azerbaijan Academy of Sciences, Baku; Azerbaijan.\\
% ...
$^{34}$$^{(a)}$Facult\'e des Sciences Ain Chock, R\'eseau Universitaire de Physique des Hautes Energies - Universit\'e Hassan II, Casablanca;$^{(b)}$Centre National de l'Energie des Sciences Techniques Nucleaires (CNESTEN), Rabat;$^{(c)}$Facult\'e des Sciences Semlalia, Universit\'e Cadi Ayyad, LPHEA-Marrakech;$^{(d)}$Facult\'e des Sciences, Universit\'e Mohamed Premier and LPTPM, Oujda;$^{(e)}$Facult\'e des sciences, Universit\'e Mohammed V, Rabat; Morocco.\\
% ...
$^{39}$Niels Bohr Institute, University of Copenhagen, Copenhagen; Denmark.\\
% ...
$^{91}$Department of Physics, Royal Holloway University of London, Egham; United Kingdom.\\
% ...
$^{99}$CPPM, Aix-Marseille Universit\'e, CNRS/IN2P3, Marseille; France.\\
% ...
$^{124}$Homer L. Dodge Department of Physics and Astronomy, University of Oklahoma, Norman OK; United States of America.\\
% ...
$^{128}$LAL, Universit\'e Paris-Sud, CNRS/IN2P3, Universit\'e Paris-Saclay, Orsay; France.\\
% ...
$^{153}$Department of Physics and Astronomy, University of Sussex, Brighton; United Kingdom.\\
% ...
$^{158}$Raymond and Beverly Sackler School of Physics and Astronomy, Tel Aviv University, Tel Aviv; Israel.\\
% ...
$^{164}$Department of Physics, University of Toronto, Toronto ON; Canada.\\
% ...
$^{a}$ Also at Borough of Manhattan Community College, City University of New York, NY; United States of America.\\
$^{b}$ Also at Centre for High Performance Computing, CSIR Campus, Rosebank, Cape Town; South Africa.\\
$^{c}$ Also at CERN, Geneva; Switzerland.\\
$^{d}$ Also at CPPM, Aix-Marseille Universit\'e, CNRS/IN2P3, Marseille; France.\\
$^{e}$ Also at D\'epartement de Physique Nucl\'eaire et Corpusculaire, Universit\'e de Gen\`eve, Gen\`eve; Switzerland.\\
% ...
$^{*}$ Deceased

\end{flushleft}


\end{document}